\documentclass[12pt]{article}
\newlength{\pubnumber} 

\catcode`\@=11
\@addtoreset{equation}{section}

\def\section{\@startsection{section}{1}{\z@}{3.5ex plus 1ex minus .2ex}
 {2.3ex plus .2ex}{\large\bf}}
\def\subsection{\@startsection{subsection}{2}{\z@}{2.3ex plus .2ex}
 {2.3ex plus .2ex}{\bf}}

\usepackage{latexsym}
\usepackage{subfigure}

\begin{document}

\begin{titlepage}
\samepage{
\setcounter{page}{0}
\rightline{December 2011}
\vfill
\begin{center}
    {\Large \bf Kaluza-Klein Masses and Couplings:\\ 
   Radiative Corrections to Tree-Level Relations\\}
\vfill
   {\large Sky Bauman$^{1,2}$\footnote{E-mail address:
      sbauman@physics.wisc.edu} $\,$and$\,$ Keith
      R. Dienes$^{3,4,2}$\footnote{E-mail address:
      dienes@physics.arizona.edu} \\}
\vspace{.10in}
   {\it 
     $^1$ Department of Physics, University of Wisconsin, Madison, WI  53706  USA\\ 
     $^2$ Department of Physics, University of Arizona, Tucson, AZ  85721  USA \\  
     $^3$ Physics Division, National Science Foundation, Arlington, VA  22230  USA\\
     $^4$ Department of Physics, University of Maryland, College Park, MD  20742  USA \\  }  
\end{center}
\vfill
\begin{abstract}
  {\rm  The most direct experimental signature of a compactified extra
   dimension is the appearance of towers of Kaluza-Klein
   particles obeying specific mass and coupling relations.  
   However, such masses and couplings are subject to
   radiative corrections.  In this paper, using techniques developed
   in previous work, we investigate the extent to which such radiative
   corrections deform the expected tree-level relations between
   Kaluza-Klein masses and couplings.  As toy models for our analysis,
   we investigate a flat five-dimensional scalar $\lambda \phi^4$ model and
   a flat five-dimensional Yukawa model involving both scalars and
   fermions.  In each case, we identify the conditions under which the
   tree-level relations are stable to one-loop order, and the
   situations in which radiative corrections modify the algebraic
   forms of these relations.  Such corrections to Kaluza-Klein spectra
   therefore have the potential to distort 
   the apparent geometry of a large extra dimension.}
\end{abstract}
\vfill
\smallskip}
\end{titlepage}

\setcounter{footnote}{0}

\def\beq{\begin{equation}}
\def\eeq{\end{equation}}
\def\beqn{\begin{eqnarray}}
\def\eeqn{\end{eqnarray}}
\def\half{{\textstyle{1\over 2}}}
\def\quarter{{\textstyle{1\over 4}}}

\def\calO{{\cal O}}
\def\calE{{\cal E}}
\def\calT{{\cal T}}
\def\calM{{\cal M}}
\def\calF{{\cal F}}
\def\calS{{\cal S}}
\def\calY{{\cal Y}}
\def\calV{{\cal V}}
\def\ibar{{\overline{\imath}}}
\def\chibar{{\overline{\chi}}}
\def\ttwo{{\vartheta_2}}
\def\tthree{{\vartheta_3}}
\def\tfour{{\vartheta_4}}
\def\ttwob{{\overline{\vartheta}_2}}
\def\tthreeb{{\overline{\vartheta}_3}}
\def\tfourb{{\overline{\vartheta}_4}}

\def\qbar{{\overline{q}}}
\def\mm{{\tilde m}}
\def\nn{{\tilde n}}
\def\rep#1{{\bf {#1}}}
\def\ie{{\it i.e.}\/}
\def\eg{{\it e.g.}\/}

\newcommand{\newc}{\newcommand}
\newc{\gsim}{\lower.7ex\hbox{$\;\stackrel{\textstyle>}{\sim}\;$}}
\newc{\lsim}{\lower.7ex\hbox{$\;\stackrel{\textstyle<}{\sim}\;$}}

\hyphenation{su-per-sym-met-ric non-su-per-sym-met-ric}
\hyphenation{space-time-super-sym-met-ric}
\hyphenation{mod-u-lar mod-u-lar--in-var-i-ant}


\def\inbar{\,\vrule height1.5ex width.4pt depth0pt}

\def\IC{\relax\hbox{$\inbar\kern-.3em{\rm C}$}}
\def\IQ{\relax\hbox{$\inbar\kern-.3em{\rm Q}$}}
\def\IR{\relax{\rm I\kern-.18em R}}
 \font\cmss=cmss10 \font\cmsss=cmss10 at 7pt
\def\IZ{\relax\ifmmode\mathchoice
 {\hbox{\cmss Z\kern-.4em Z}}{\hbox{\cmss Z\kern-.4em Z}}
 {\lower.9pt\hbox{\cmsss Z\kern-.4em Z}} {\lower1.2pt\hbox{\cmsss
 Z\kern-.4em Z}}\else{\cmss Z\kern-.4em Z}\fi}

\long\def\@caption#1[#2]#3{\par\addcontentsline{\csname
  ext@#1\endcsname}{#1}{\protect\numberline{\csname
  the#1\endcsname}{\ignorespaces #2}}\begingroup \small
  \@parboxrestore \@makecaption{\csname
  fnum@#1\endcsname}{\ignorespaces #3}\par \endgroup}
\catcode`@=12

\input epsf

\section{Introduction
\label{intro}
}
\setcounter{footnote}{0}

The existence of Kaluza-Klein (KK) states is perhaps the most
important phenomenological prediction of extra dimensions, and any
future search for physics beyond the Standard Model will involve a hunt
for signs of these particles.  For this reason, it is vital to understand
the properties of these states and the effects that they induce
on low-energy physics. 
Of course,
one important way in which excited KK states can affect low-energy physics
is through the radiative corrections that they induce for zero-mode masses and couplings.
Indeed, over the past decade, 
a significant body of literature has developed in which this topic is studied
in a variety of contexts and from a variety of perspectives.

However, 
with only a few exceptions,
relatively little attention has been paid to the radiative effects 
that the excited KK states may have {\it on their own masses and couplings}\/. 
Since these excited KK states are likely to be our only direct experimental probes into
the apparent geometry of the compactification manifold, it is
important to understand the extent to which such radiative corrections can
distort the expected tree-level relations that the KK masses and
couplings can be expected to satisfy, and which would ultimately be used as
evidence of a geometric underpinning for such states.

To help sharpen the discussion, let us consider the simplest possible case
of a single extra dimension compactified on a circle.  At tree level, the masses of
the corresponding KK states can be expected to obey the well-known
dispersion relation
\beq
     m_{n}^{2} ~=~ m^2 + \frac{n^2}{R^2}~
\label{kaluza2}
\eeq
where $m_n$ is the mass of the $n^{\rm th}$ KK mode, where
$m$ is the ``bare'' mass associated with our original five-dimensional field,
and where $R$ is the radius of the extra dimension.  
Note that this result
assumes only that the extra dimension is flat and that the original
theory obeys five-dimensional (5D) Lorentz symmetry.  Likewise, at
tree level, the couplings in a Lorentz-invariant theory on an extra
dimension are universal, independent of mode number.  Specifically, if
$\lambda_{n,n',...}$  represents a tree-level coupling between KK modes $(n, n',...)$,  
then
\beq
    \lambda_{n,n',...} ~=~ \lambda\,\delta_{n+n'+...,0}~
\label{uni}
\eeq
where $\lambda$ is a constant related to the five-dimensional ``bare'' coupling
and where the delta-function 
enforces 5D momentum conservation at the associated vertex.  The important point is
that $\lambda_{n,n',...}$ takes this highly restricted form, depending
on the KK mode numbers $(n,n',n'',...)$ only insofar as they determine 
whether the coupling vanishes or
takes a fixed, mode-independent value. 

Like the masses and couplings in any theory, however, the masses and
couplings of KK states can receive radiative corrections.  Thus, it is
possible that the tree-level relations in Eqs.~(\ref{kaluza2})
and~(\ref{uni}) will no longer hold once these masses and couplings
are replaced by their one-loop renormalized values.
At first glance, it might seem that the forms of Eqs.~(\ref{kaluza2})
and~(\ref{uni}) are fixed by 5D Lorentz invariance.
However, we must recall that 5D Lorentz invariance is actually broken by the
compactification from five to four dimensions.
The effects of this compactification 
are what allow more complicated dispersion relations to emerge
in the fully quantum-mechanical theory.

Focusing specifically on the mass relation in Eq.~(\ref{kaluza2}), we can imagine a
number of potential outcomes depending on the specific theory in question.  
One possibility is that the one-loop
renormalized KK masses will continue to obey  a relation that preserves the form
of Eq.~(\ref{kaluza2}) --- \ie, that all radiative corrections can be bundled
into a new effective bare mass $m$ or a new effective radius $R$.
Despite the fact that $m$ and $R$ are merely fixed parameters
describing our ultraviolet theory, we shall refer to these outcomes as
effective ``renormalizations'' of these quantities.  However, the breaking of
5D Lorentz invariance might also allow the spectrum of
renormalized KK masses  to have an entirely new dependence on
mode number, implying that even the {\it forms}\/ of  the tree-level
relations might be violated.

More precisely, we can classify the different types of quantum
corrections that our squared KK masses may experience:
\begin{itemize}
\item {\sl Case \#1:}  The corrections to each $m_n^2$ are independent 
     of mode number $n$. In this case, 
     the bare mass $m$ is effectively renormalized,  but the KK
     dispersion relation retains the same mathematical form as it had
     at tree level.  In
     this case, 5D Lorentz invariance is preserved locally.  However,
     since compactification breaks 5D Lorentz invariance globally by
     singling out the compactified extra dimension, this occurrence would be
     entirely unexpected.  We shall nevertheless give an example where this
     phenomenon arises to one-loop order in Sect.~3.
\item {\sl Case \#2:}  The corrections to $m_n^2$ are proportional to the square of 
      the mode number $n$.  
     In this case, we can bundle the renormalizations into an
     effective rescaling or renormalization of the radius $R$.  To the extent that the
     radius is an arbitrary parameter and the form of the general KK
     mass relation is preserved, this also would not indicate a direct
     local breaking of 5D Lorentz invariance.   
     As such, this case would also be unexpected, 
      just like Case~\#1 above.
     However, since such radiative corrections would
     manifest themselves as effectively modifying the value of $R$,  
      it would {\it appear}\/ that our underlying compactification geometry is distorted somewhat, with
      the radius of the circle shifting slightly.
     We stress, however, that    
     this is not an actual geometric effect since the underlying compactification geometry 
        is presumably unchanged (unless there are also renormalizations of the higher-dimensional metric).
     This is therefore merely a change in the {\it apparent}\/ compactification geometry, as 
         inferred through the masses of KK states.
\item {\sl Case \#3:}  The corrections to $m_n^2$ depend on mode number $n$ non-quadratically. 
      In this case, it turns out that there is a particularly relevant division into two sub-cases
     which we shall consider:
\begin{itemize}
\item {\sl Case \#3a:}  The masses of an infinite subset of states in the KK tower shift 
           according to Case~\#1 or Case~\#2
            (corresponding to shifts in the values of $m$ or $R$),
             but this is not true of the entire KK tower.
            Thus, the KK dispersion relation is broken 
            for the KK tower on a whole. We shall refer to this as an {\it implicit}\/ 
           violation of the KK dispersion relation.
\item {\sl Case \#3b:}   The KK dispersion relation does not survive for 
           any infinite subset of states in the KK tower.  
            We shall refer to this as an {\it explicit}\/ 
           violation of the KK dispersion relation.
\end{itemize}
In this paper, we shall see explicit examples of both of these cases.
Note that for either of these two sub-cases,  
the KK masses as a whole no longer obey Eq.~(\ref{kaluza2}).
It would therefore seem that these KK states could no longer be identified
as the Kaluza-Klein excitations 
of a quantum field compactified on a circle ---
\ie, the apparent compactification geometry of the extra dimension 
would appear distorted in such a way in such cases that not even an 
underlying circle is recognizable. 
\end{itemize}

In this paper, our goal is to begin to develop an understanding of 
the sorts of theories which might lead to corrections in each class.
Towards this end, we shall therefore study two ``toy'' models:
$\phi^4$ theory and Yukawa
theory, each in five dimensions with a single extra dimension compactified on a circle. 
For each of these two theories, we shall obtain results for
the radiative corrections to the masses and couplings 
of the KK modes, and examine the properties of the physics which results.

Both of these toy models may ultimately be relevant to
the Higgs sector of the 5D Standard Model. 
Despite this fact, we emphasize that the primary purpose of this paper is not phenomenological,
and indeed many of these radiative corrections will turn out to be numerically fairly small. 
Rather, our primary focus will be on the mathematical forms of the radiative
corrections that emerge in each case, and on the general mathematical patterns that
describe the deformations of KK masses and couplings which emerge as
a result of radiative corrections.
For example, one unexpected result we shall find is that the masses of the fermions in
    the Yukawa theory receive corrections that actually grow with mode number.
    Another is that a $\gamma^5$ interaction is radiatively induced in this theory.
Even the $\phi^4$ theory will hold some surprises.  For example, as we shall demonstrate,
    radiative corrections tend to enhance 
   the couplings involving the production of excited KK modes.

This paper is organized as follows.
First, in Sect.~2, we begin with some general comments concerning 
renormalization and regulators in ``mixed'' spacetimes in which some dimensions
are compactified and others are not.  We also describe
the general setup we shall be employing.
Then, in Sect.~3, we analyze the $\lambda \phi^4$ theory,
concentrating on corrections to the masses and couplings of the KK states.
In Sect.~4, we then proceed to consider the Yukawa theory;
as we shall see, the Yukawa theory is significantly more complex than the
$\lambda \phi^4$ theory due to the involvement of fermions and issues
of parity and chirality.
Finally, in Sect.~5, we present our conclusions and discuss
how our results connect with other calculations which have 
previously appeared in the literature.

\section{General setup}
\setcounter{footnote}{0}

As stated in the Introduction, our goal is to determine how the
tree-level masses and couplings of KK modes behave under
renormalization.   Before proceeding to examine the cases of specific
toy models, however, there are some general remarks which are in order
and which will apply to all cases we shall consider.

In general, KK masses and couplings will accrue radiative corrections which
are divergent.
However, although each of these corrections is individually divergent,
the difference between a correction corresponding to an excited KK mode 
and that corresponding to the zero mode
is observable and therefore finite~\cite{1a,1b}.
For example, although the mass of the zero mode and mass of the first excited mode 
of a KK tower will each generally accrue radiative corrections  
which are infinite, the {\it difference}\/ between these masses (\ie, the
mass splitting between these KK modes) is expected to remain
finite even after renormalization. 
The first step in determining such radiative corrections is therefore to recast
equations such as Eq.~(\ref{kaluza2})
into forms whose corrections will be nothing other than these finite differences.
In other words, we wish to express these tree-level equations 
as relations directly between
measurable, four-dimensional quantities, eliminating the bare
Lagrangian parameters $m$ and $\lambda$ in the process.

In the case of Eq.~(\ref{kaluza2}), this is not hard to do.
Since it follows from Eq.~(\ref{kaluza2}) that $m_0=m$ at tree level,
we can rewrite Eq.~(\ref{kaluza2}) in the tree-level form
\beq
          m_n^2 ~=~ m_0^2 + {n^2\over R^2}~,
\label{kaluza3}
\eeq
whereupon it follows that any possible one-loop radiative correction to this result
must be finite and take the form
\beq
                   m_{n}^2 ~=~ m_0^2 + {n^2\over R^2} +
               \frac{X_{n}^{(m^2)}}{R^2} 
\label{firstcorr}
\eeq
where 
$X_n^{(m^2)}$ represents the finite mass correction term.
Note that we have chosen to explicitly scale out a factor of $R^2$ in this correction
term so that the quantity $X_n^{(m^2)}$ is dimensionless.

Given this definition for $X_n^{(m^2)}$ as a relative mass correction,
we see that Case~\#1 from the Introduction corresponds to
$X_n^{(m^2)}=0$,
while Case~\#2 corresponds to 
$X_n^{(m^2)}\sim n^2$.
By contrast, $X_n = {\rm constant}\not= 0$ is actually an example
of Case~\#3a, since this corresponds to a situation
in which all of the excited states in our KK tower have a uniform shift relative
to the zero mode.  Thus, although the infinite tower of {\it excited}\/ states by themselves behave
according to Case~\#1, the entire KK tower (including the zero mode) does not.

In a similar way, we may also recast Eq.~(\ref{uni}) in the form
\beq
    \lambda_{n,n',...} ~=~ \lambda_{0,0,...} \,\delta_{n+n'+...,0}~,
\label{uni2}
\eeq
whereupon a corresponding one-loop equation should take the form 
\beq
         \lambda_{n,n',...} ~=~ 
         \left[ \lambda_{0,0,...} + 
          X_{n,n',...}^{(\lambda)}\right]~ \delta_{n+n'+...,0}~.
\label{secondcorr}
\eeq
where 
$X_{n,n',...}^{(\lambda)}$ is likewise a finite coupling correction.

The goal of this paper
is to calculate these finite corrections $X^{(m^2)}_{n}$ and $X^{(\lambda)}_{n,n',...}$
to one-loop order in two different theories, and to explore the 
properties of these corrections.
Of course, the emergence of such correction terms ultimately reflects
the breaking of the higher-dimensional Lorentz invariance that is
induced by the compactification of the fifth dimension on a circle.  
We remark, however, that although
compactification of the fifth dimension breaks 5D Lorentz invariance,
translational invariance along the fifth dimension (and thus
conservation of the corresponding momenta)
is still maintained.
It is for this reason that our radiatively corrected couplings
$\lambda_{n,n',...}$ must still be proportional to an overall
Kronecker $\delta$-factor, as indicated in Eq.~(\ref{secondcorr}).

At first glance, given a specific theory,
it might seem to be a rather straightforward exercise
to evaluate the radiative corrections $X$ in
Eqs.~(\ref{firstcorr}) and~(\ref{secondcorr}).  However, as discussed
in Refs.~\cite{1a,1b}, there are numerous subtleties which come into
play.  The chief complication is that although we expect our
calculations to result in finite relative corrections $X$, the
correction to each individual KK mass and coupling will itself
be infinite, and we must therefore utilize a particular regulator scheme
in order to extract meaningful results.  However, in so doing,
it is critical that we choose a regulator which preserves not only the
four-dimensional Lorentz invariance that remains after the 
compactification, {\it but also the original
higher-dimensional Lorentz invariance which existed prior to
compactification.}\/  This is because a regulator must, by design,
be capable of handling ultraviolet (\ie, local short-distance) divergences, and 
the physics of the ultraviolet limit is governed by
such {\it five-dimensional}\/ symmetries 
in which the global process of compactification plays no role.
Moreover, in contexts in which our original
higher-dimensional Lagrangian contains a gauge symmetry, our regulator
should respect this higher-dimensional gauge invariance as well.

This is an important point.
Indeed, use of any regulator which fails to respect
the approprite five-dimensional UV symmetries 
such as 5D Lorentz invariance
would introduce
spurious, unphysical 5D Lorentz-violating contributions into the $X$
corrections, and it would be difficult to disentangle these spurious contributions
from the {\it bona-fide}\/ 
physical effects of the 5D Lorentz violation induced by
compactification.  
This would be completely analogous to calculating a one-loop correction
to the photon mass in QED with a regulator that breaks gauge invariance:
a non-zero result will generically arise, but this would merely be an artifact of the calculational technique
and would not reflect the true underlying physics.
Our current situation with 5D Lorentz invariance is similar, except
that the compactification itself also induces a breaking of 5D Lorentz symmetry. 
However, our goal is to 
study the effects of this compactification (as manifested by the appearance of radiative corrections $X$
that deform the forms of the tree-level KK mass and coupling relations)
without mixing such effects with the unphysical
effects of having chosen an unsuitable regulator.  

In Refs.~\cite{1a,1b}, two regulators were developed that can handle precisely
such calculations.  
These are the so-called ``extended hard cutoff'' (EHC)
regulator scheme and the so-called ``extended dimensional regularization'' (EDR)
scheme.  
Although based on traditional four-dimensional regulators,
the key new feature of these higher-dimensional regulators is that
they are specifically designed to handle mixed spacetimes in which some
dimensions are infinitely large  and others are compactified.
Moreover, unlike most other regulators which have been used in
the extra-dimension literature, these regulators are designed to
respect the original higher-dimensional Lorentz symmetries that exist prior to
        compactification, and not merely the four-dimensional symmetries which
        remain afterward.  As we have discussed above,
this distinction is particularly relevant for
        calculations of the physics of the excited Kaluza-Klein modes
        themselves, and not merely their radiative effects on zero modes.  By
        respecting the full higher-dimensional symmetries, our regulators
        avoid the introduction of spurious terms which would not have been
        easy to disentangle from the physical effects of compactification.

Using the regulators developed in Refs.~\cite{1a,1b}, 
we can evaluate the corrections $X$ to one-loop order in a variety of different theories.
Regardless of the theory, however, it turns out~\cite{1a,1b}
that one-loop radiative corrections $X_n$  with a single KK index $n$ can 
generally be expressed in the form
\beq
         X_n ~=~  \sum_{r = -\infty}^{\infty}\frac{1}{|n|}\sum_{j = 0}^{|n| - 1}
            \int_{0}^{1}dv  \left[ \alpha_n(r,v,j) - \alpha_{0}(r,v) \right]
\label{Xform}
\eeq
where $\alpha_0$ and $\alpha_n$ are {\it finite,
regulator-independent}\/ functions and where the summations and
integration in Eq.~(\ref{Xform}) are all {\it absolutely convergent}\/.  
Here $v$ is a Feynman parameter, and it is assumed that all of the relevant
diagrams involved in such radiative corrections can be evaluated with
the use of a single Feynman parameter.  Indeed, an expression analogous to
Eq.~(\ref{Xform}) is available in certain cases requiring multiple Feynman
parameters~\cite{1a}, and we shall see an example of this in Sect.~4. 

At first glance, the result in Eq.~(\ref{Xform}) might not
seem particularly noteworthy.  After all,
an expression of this general form 
arises immediately upon a straightforward 
application of the Feynman rules, with
appropriate one-loop integrals taking the 
place of the $\alpha$-functions in Eq.~(\ref{Xform}).
However, such integrals are generally divergent.  
The important point in Eq.~(\ref{Xform}), by contrast,
is that the $\alpha$-functions in Eq.~(\ref{Xform})
are both {\it finite}\/ and {\it regulator-independent}\/;
moreover, 
with the appropriate 
$\alpha$-functions inserted into Eq.~(\ref{Xform}),
it turns out that the summations and integration in Eq.~(\ref{Xform}) are also convergent.
That such $\alpha$-functions
exist is the main substance of the results of Refs.~\cite{1a,1b},
and it is the use of the special regulators in Refs.~\cite{1a,1b} which allows
these functions to be obtained.
The explicit forms of these $\alpha$-functions therefore
encapsulate the physical effects of the one-loop renormalizations without
including any of the spurious mathematical artifacts that might arise due to the use
of regulators which do not respect the full ultraviolet symmetries of the problem.

In this paper, therefore, we shall assume that the reader is familiar with 
the calculational techniques leading to these $\alpha$-functions, 
and we shall simply
quote our final results for the specific theories at hand.
We also note that in this paper we will calculate radiative corrections
to KK {\it couplings}\/ as functions of a canonical (non-Wilsonian) renormalization scale $\mu$.
By contrast, we will calculate radiative corrections to KK {\it masses}\/ on resonance
(\ie, with mass renormalization conditions imposed on shell).

\section{$\lambda \phi^4$ theory
\label{summary}
}
\setcounter{footnote}{0}

As our first simple toy model, in this section we will  examine the
case of a purely bosonic $\lambda\phi^4$ theory on a circular extra
dimension of radius $R$.

We begin with a five-dimensional theory defined by the $\phi^4$ action
\begin{equation}
    S ~=~ \int d^4 x \int_{0}^{2\pi R}dy \, \left\lbrack
    \frac{1}{2}\partial^M \phi^* \partial_M \phi -  \frac{1}{2}m^2
    \phi^* \phi -  \frac{\lambda^{(5)}}{4!}\phi^4 \right\rbrack~ ,
\label{phaction}
\end{equation}
where $y$ is the coordinate along the extra dimension,  where
$x^M\equiv (x^\mu,y)$,  where $\lambda^{(5)}$ is a 5D
coupling,  and $\phi$ is assumed to be a real scalar.  Proceeding in
the usual way,  we decompose the $\phi$-field  in terms of
Kaluza-Klein modes
\beq
     \phi (x^{\mu},y) ~=~ \frac{1}{\sqrt{2\pi R}}\,\sum_{n\in\IZ}
     \,\phi_n (x^{\mu})\, e^{iny/R}~,
\label{kkph}
\eeq
and substitute this back into the original action in
Eq.~(\ref{phaction}). Since the 5D field $\phi$ is real,
we have $\phi^\ast_n=\phi_{-n}$.
Integrating over $y$, we thus obtain a purely four-dimensional action
of the form
\beq
    S ~=~ \int d^4 x \,  \biggl( \half \, \sum_n \, \partial^{\mu}
    \phi^{*}_n \partial_{\mu} \phi_n  - \half \, \sum_n \, m_{n}^2
    \phi^{*}_n \phi_n  
       - {1\over 4!}\,
    \sum_{n_i\in\IZ} \lambda_{n_1,n_2,n_3,n_4}
    ~\phi_{n_4}\phi_{n_3}\phi_{n_2}\phi_{n_1} \biggr)~
\label{kk4}
\eeq
where the 4D KK masses $m_n^2$ are given exactly as in Eq.~(\ref{kaluza2})
and where the 4D couplings $\lambda_{n_1,n_2,n_3,n_4}$ are given by  
a special case of Eq.~(\ref{uni}):
\beq
         \lambda_{n_1,n_2,n_3,n_4} ~=~   \frac{\lambda^{(5)}}{2\pi R}~
            \delta_{n_1+n_2+n_3+n_4,0}~.
\label{pre4Dlam}
\eeq
As discussed above, the $\delta$-function in Eq.~(\ref{pre4Dlam}) 
expresses the conservation of five-momentum at a vertex, as
appropriate for  compactification on a circle in which translational
invariance in the extra dimension is preserved.  

\begin{figure}[t]
\centerline{
   \epsfxsize 5.8 truein \epsfbox {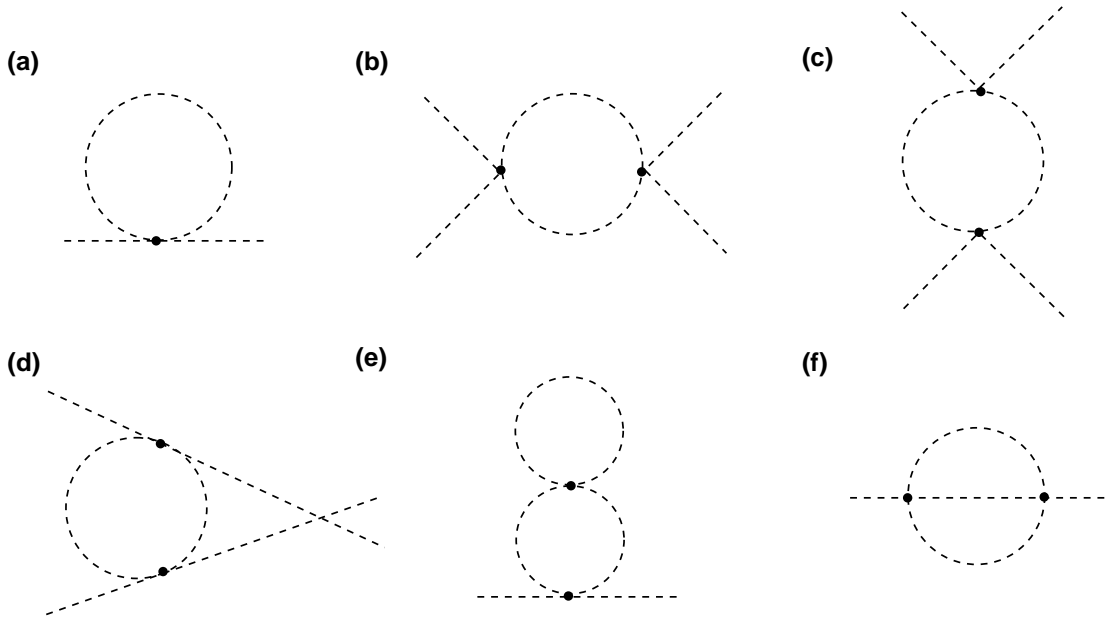} }
\caption{Relevant diagrams for renormalization in the $\lambda\phi^4$ theory:
  (a) one-loop mass renormalization; (b,c,d) one-loop coupling
  renormalizations; (e,f) two-loop mass renormalizations.  Of the
  diagrams contributing to mass renormalization, only diagram~(f)
  yields a contribution which depends on the Kaluza-Klein mode number
  of the external particle.  Thus, only diagram~(f) produces
  Lorentz-violating deformations away from the form of the tree-level
  Kaluza-Klein mass relation.}
\label{fig1}
\end{figure}

Following the steps outlined in Sect.~2, we can now convert
these mass and coupling relations to the forms given in
Eqs.~(\ref{kaluza3}) and (\ref{uni2}), recasting them as direct tree-level relations
between observable, four-dimensional quantities.
We therefore 
expect that these equations will accrue finite one-loop corrections of the
forms given in Eqs.~(\ref{firstcorr}) and (\ref{secondcorr}).
In order to explicitly calculate these radiative corrections $X_n^{(m^2)}$ and
$X^{(\lambda)}_{n_1,n_2,n_3,n_4}$ in the current $\lambda \phi^4$ theory,
we must evaluate the one-loop diagrams those shown in Fig.~\ref{fig1}.
Using the regulators developed in Refs.~\cite{1a,1b}, 
we then find the following results.

\subsection{Mass corrections}

We first examine the mass corrections $X^{(m^2)}_n$ in this theory.
Recall from Eq.~(\ref{Xform}) that each correction $X_n$ can
be expressed in terms of corresponding functions $\alpha_n$ and $\alpha_0$.
However, to one-loop order, it turns out that
\beq
    X^{(m^2)}:~~~~~~~~          \alpha_n~=~ \alpha_0~=~0~~~~~  
                  {\rm for~all}~n~.
\eeq
In other words, the corresponding mass corrections $X^{(m^2)}_n$ all
vanish, and the tree-level mass relation in Eq.~(\ref{kaluza3})
remains intact to one-loop order.

This is clearly an example of Case~\#1 from the Introduction.  We
emphasize that this does {\it not}\/ mean that there are no radiative
corrections to the individual KK masses --- indeed, each individual KK
mass receives a correction which is infinite.  However, these mass
corrections are all equal to each other.   This implies that the
corrections to each KK mass are independent of the mode number $n$,
and consequently can be bundled within $m_0$.  Equivalently, these
radiative corrections can be absorbed within a single shift in the
bare parameter $m$ in our original higher-dimensional Lagrangian.
Thus the {\it relation}\/ between zero-mode masses and excited KK
masses remains unchanged.

It is easy to see why this situation arises for the $\lambda \phi^4$
theory.  The relevant diagram for one-loop mass renormalization is
shown in Fig.~\ref{fig1}(a).  Because of the topology of this diagram,
the momentum that flows through the loop is wholly independent of the
Kaluza-Klein index on the  external line.  Thus, each external
Kaluza-Klein state accrues exactly the same mass correction, and it is
possible to  bundle this into an effective ``renormalization'' of the
constant term $m_0^2$.  In other words, only
one mass counterterm is needed, and the KK mass relations predicted by
5D Lorentz invariance are preserved.

We stress, however, that this is merely a one-loop phenomenon.  
For example,
two-loop diagrams contributing to mass renormalization are
shown in  Figs.~\ref{fig1}(e) and \ref{fig1}(f).  While
Fig.~\ref{fig1}(e) also leads to a mass renormalization which is
independent of the Kaluza-Klein number of the external line, the
contribution from Fig.~\ref{fig1}(f)  clearly depends non-trivially on
this index.  Thus, to two-loop order, Fig.~\ref{fig1}(f) represents
the only diagram leading to radiative effects which break the tree-level
mass relations.

\subsection{Coupling corrections}

We now turn to the coupling corrections $X^{(\lambda)}_{n_1,n_2,...}$ in $\lambda \phi^4$
theory.  It is
here that violations of 5D Lorentz invariance will appear at one-loop
order.

\begin{figure}[t!]
\centerline{
   \epsfxsize 3.5 truein \epsfbox {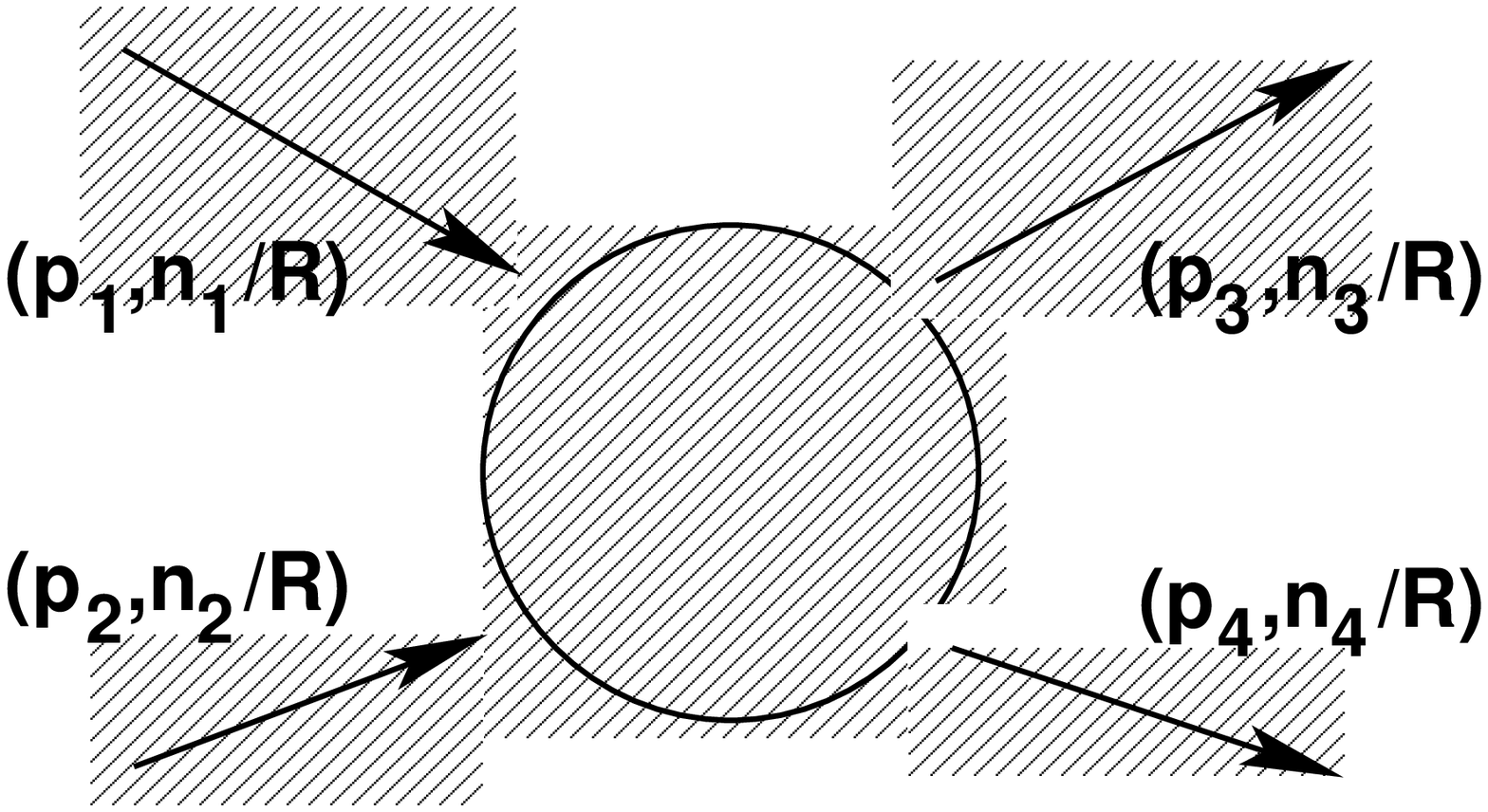} }
\caption{Momentum labeling for 5D Mandelstam variables in Eq.~(\ref{mandel}).}
\label{figblob}
\end{figure}

The one-loop diagrams which contribute to the  radiative corrections
to the four-scalar couplings are shown in Figs.~\ref{fig1}(b),
\ref{fig1}(c), and \ref{fig1}(d).  These are respectively $s$-, $t$-,
and $u$-channel diagrams, and as such  they can be treated similarly.
If we establish our  momentum-labeling conventions for incoming and
outgoing states  as indicated in Fig.~\ref{figblob}, then the
corresponding Mandelstam variables for our {\it five}\/-momenta take
the forms
\beqn
        s &=& (p_1 + p_2)^2 - (n_1 + n_2)^2 /R^2~,~\nonumber\\ 
        t &=& (p_1 - p_3)^2 - (n_1 - n_3)^2 /R^2~,\nonumber\\  
        u &=& (p_1 - p_4)^2 - (n_1 - n_4)^2 /R^2~.
\label{mandel}
\eeqn
As customary in four-dimensional theories, these variables continue to satisfy the on-shell 
relation $s+t+u=4m^2$, where $m$ is now the {\it five-dimensional}\/ mass given in Eq.~(\ref{phaction}).

We then find that  at one-loop order, the couplings
$\lambda_{n_1,n_2,n_3,n_4}$ are no longer universal;  new corrections
$X^{(\lambda)}_{n_1,n_2,n_3,n_4}$ are introduced.  Defining these
corrections through the relation
\beq
    \lambda_{n_1,n_2,n_3,n_4} =  \left[ \lambda_{0000} +
        \frac{\lambda^2}{4\pi} X^{(\lambda)}_{n_1,n_2,n_3,n_4} \right]\,
        \delta_{n_1+n_2-n_3-n_4,0 }~,
\eeq
we find that they each receive three contributions:
\beq
    X^{(\lambda)}_{n_1,n_2,n_3,n_4} ~=~ \xi_{n_1 + n_2}(s) 
               + \xi_{n_1 - n_3}(t) + \xi_{n_1 - n_4}(u)~.
\label{lcoup}
\eeq
These three contributions correspond to the diagrams in
Figs.~\ref{fig1}(b), \ref{fig1}(c), and \ref{fig1}(d) respectively.
Unlike $X^{(\lambda)}$ itself, the $\xi$-functions depend on only a
single KK index and a single Mandelstam variable;  they can thus be
expressed in the form in Eq.~(\ref{Xform}).  Using the techniques
discussed in Refs.~\cite{1a,1b}, we then find that the  corresponding
$\alpha$-functions are given by
\beq
\xi_n (s):~~~~~~~~ \cases{
   {\displaystyle \alpha_{n}(r,v,j;s) ~=~\frac{1}{4\pi}\log[(r-v)^2 +
      \mathcal{M}^2 ((v+j)/|n|;s)R^2]}~ & \cr ~ & \cr {\displaystyle
      \alpha_{0}(r,v;s)~=~   \frac{1}{4\pi}\log[r^2 + \mathcal{M}^2
      (v;s)R^2]}~ & \cr}
\label{x_sum}
\eeq
where
\beq
          \mathcal{M}^2 (x;s) ~\equiv~ x(x - 1)s + m^2~. 
\eeq
For notational simplicity throughout the rest of this paper, we shall henceforth define
\beq
        y_n ~\equiv~ \cases{ v & for $n=0$ \cr
                          (v+j)/|n| & for $n\not=0$\cr}
\label{ydef}
\eeq
and 
\beq
        \rho_n ~\equiv~ \cases{ r & for $n=0$ \cr
                          r-v & for $n\not=0$~.\cr}
\label{rhodef}
\eeq
We can then simply write our result in the compact form
\beq
\alpha_n~=~ \frac{1}{4\pi}\log[\rho_n^2 + \mathcal{M}^2 (y_n;s)R^2]~.
\label{x_sum_rho}
\eeq

These results are completely general.
However, in order to evaluate these results numerically, it is necessary to choose
specific values for the kinematic Mandelstam variables $(s,t,u)$.
At first glance, one might be tempted to impose the sorts of
renormalization conditions that would apply to processes involving
only zero-mode fields, such as $s=4m^2$ and $t=u=0$.
However, such conditions correspond to situations in which all of the
modes have vanishing spatial momenta, and thus cannot accommodate the
sorts of processes
which are of interest to us, such as those involving
the production of excited KK modes.
Similarly, one might consider a renormalization condition such
as $s=t=u= -\mu^2$, where $\mu$ is the floating energy scale associated
with an experiment.  However, these conditions cannot be satisfied when any of the
incoming or outgoing particles are on shell.

We shall therefore adopt renormalization conditions
of the form
\beq
              s ~=~ \mu^2 + 4m^2~,~~~~~~ t ~=~ u ~=~ -\mu^2/2~.
\label{mudef}
\eeq
Note that in the center-of-mass frame (defined as that
frame in which all spatial components of the total five-momentum of the system vanish),
we may identify the energy scale $\mu$ as
\beq
          \mu^2 ~=~ 4 (\vec p^{\,2} + p_5^2) 
\eeq
where $\vec p$ and $p_5$ are the spatial momentum components of any single particle alone. 
(Of course, in this center-of-mass frame, the assigned KK mode-numbers $n\sim R p_5$ of these states might differ
from those we have been assigning in our four-dimensional ``lab'' frame.)
However, despite the somewhat intuitive form of the renormalization conditions in Eq.~(\ref{mudef}), it is 
important to realize that these conditions  
place special restrictions on the scattering angle.
Such restrictions are unfortunately unavoidable, and will arise for
 {\it any}\/ such constraint on the three Mandelstam variables.

\begin{figure}[ht]
\centerline{
   \epsfxsize 3.5 truein \epsfbox {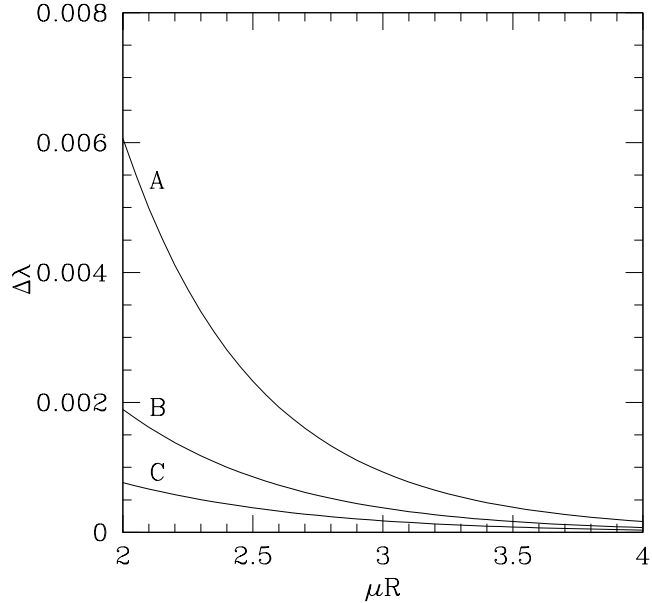} }
\caption{
   One-loop enhancement of the coupling $\lambda_{0,0,1,-1}$  for the
   production of the lowest-lying excited KK states.  We plot
   $\Delta\lambda\equiv X_{0,0,1,-1}=  (\lambda_{0,0,1,-1} - \lambda_{0000})/\chi$ as
   a function of the energy scale $\mu R$,  where $\chi\equiv
   \lambda_{0000}^2 /4\pi$.  Curves A, B, and C respectively
   represent the cases with $m^2 R^2 = \lbrace 0, 0.25, 0.5\rbrace$.
   As always, the four incoming and outgoing KK modes are taken to be
   on shell, and the scale $\mu$ is defined according to the
   conditions in Eq.~(\protect\ref{mudef}).  Note that in this plot,
   the scale $\mu$ runs from the $(n_3,n_4)=(1,-1)$ threshold energy
   at which the lowest-lying excited KK states can be produced to the
   $(n_3,n_4)=(2,-2)$ threshold energy at which the second-lowest
   excited states can be produced.  We observe that in general, these
   radiative corrections are greater for smaller five-dimensional masses $m$ and
   decrease as functions of $\mu$.}
\label{fig2b}
\end{figure}

In Fig.~\ref{fig2b}, we plot the difference between 
the one-loop $\lambda_{0,0,1,-1}$ coupling and the
one-loop $\lambda_{0000}$ coupling as 
a function of $\mu$.  This difference, of course, would have been
zero at tree level, and reflects the breaking of 5D Lorentz invariance
that appears at one-loop order in this theory.
Note that $\lambda_{0,0,1,-1}$ is the coupling which governs the process by
which two zero-mode states scatter/annihilate to produce two
lowest-lying excited KK states.  As we see from Fig.~\ref{fig2b}, one-loop
effects cause 
$\lambda_{0,0,1,-1}$ to become {\it larger}\/ than 
$\lambda_{0000}$.  This implies that there is a small
enhancement of the coupling between the zero mode  and the
first-excited KK mode relative to the couplings amongst the zero
modes themselves.  Although this enhancement is extremely small, we
see from Fig.~\ref{fig2b} that it is largest precisely  at the threshold for the
production of the first-excited mode, falling significantly as $\mu$ increases.  
We also observe that this enhancement
decreases as the five-dimensional scalar mass $m$ increases, 
and ultimately vanishes as $m\to\infty$.

It is clear from this plot that the one-loop coupling
corrections in the $\lambda \phi^4$ theory are exceedingly small.  However, 
we shall see that the analogous corrections in Yukawa theory will be significantly larger.

Finally, we observe that for certain values of $s$, the function
$\alpha_n^{(\xi)}(s)$ in Eq.~(\ref{x_sum}) can be complex.  Although
the imaginary part of an amplitude can be important,  only the real
part of an amplitude plays a role in the renormalization of Lagrangian
parameters  such as masses and couplings.   Therefore, unless
explicitly stated otherwise, it is to be understood throughout the
remainder of this paper that we are implicitly taking the real part of
any expression which describes the magnitude of a radiative correction
for any KK parameter.

\section{Yukawa theory
\label{Yuksult}}

We now turn to the case of 5D Yukawa theory in which a scalar particle
interacts with a Dirac fermion.  In some sense, this is the next-simplest 
theory to consider.  Moreover, as we shall see, the structure
of the radiative corrections is far more intricate, both for the KK
masses and for the couplings.

For Yukawa theory, we will consider two cases: one in which the scalar
is real, and the other in which it is complex.  In the case of a real
scalar, we shall take the 5D action to be
\beq
    S = \int d^4 x \int_{0}^{2\pi R}dy\left\lbrack \half \partial_M
    \phi \partial^M \phi  - \half  m_{\phi}^2 \phi^2 + \bar{\psi}(i
    \gamma^M \partial_M - m_{\psi})\psi  - G \phi \bar{\psi} \psi
    \right\rbrack
\label{Yukac}
\end{equation}
where $\phi$ and $\psi$ respectively denote the scalar and Dirac fermion
(with five-dimensional masses $m_\phi$ and $m_\psi$ respectively)
and where $G$ is the Yukawa coupling between the two.
In the case of a complex scalar, by contrast, our action is slightly
modified:
\beq
S = \int d^4 x \int_{0}^{2\pi R}dy  \left\lbrack \half \partial_M
        \phi^* \partial^M \phi - \half m_{\phi}^2 \phi^* \phi  +
        \bar{\psi}(i\gamma^M \partial_M - m_{\psi})\psi - G (\phi
        \bar{\psi}\psi  + {\rm h.c.}) \right\rbrack ~.
\label{5Dcomp}
\eeq
As we shall see, these two cases lead to somewhat different results.
Note that in both cases, our gamma-matrices take the form
$\gamma^M\equiv (\gamma^\mu,\tilde \gamma^5)$ where 
$\tilde \gamma^5 \equiv i\gamma^5 = -\gamma^0\gamma^1\gamma^2\gamma^3$. 

Performing the KK reduction of this theory is relatively straightforward.
We first consider the case in which $\phi$ is real.  The KK
decomposition of the scalar is again given by Eq.~(\ref{kkph}), while the
KK decomposition of the fermion takes the analogous form:
\beq
     \psi (x^{\mu},y) ~=~ \frac{1}{\sqrt{2\pi R}}\,\sum_{n\in\IZ} 
         \,\psi_n (x^{\mu})\, e^{iny/R}~.
\label{kkphf}
\eeq
We then obtain the effective four-dimensional action
\beqn
     S &=&  \int d^4 x \biggl\lbrack \half \sum_n \partial_\mu
           \phi_n^\ast \partial^\mu \phi_n  ~+~  \sum_n
           \overline{\psi}_n i \gamma^\mu \partial_\mu \psi_n
           ~-~\half \sum_n  m_{\phi n}^2  \phi^\ast_n
           \phi_n\nonumber\\ &&~~~~~~~ ~-~  \sum_n \overline{\psi}_n
           m_{\psi n} \psi_n  ~-~ \sum_n \sum_{n'} \, \phi_{n'-n} \,
           \overline{\psi}_{n'} \, \hat{g}_{n,n'} \, \psi_n ~+~ ...
           \biggr\rbrack~
\label{genform2}
\eeqn
where the tree-level boson masses are given by
\beq
   m_{\phi n}^2 ~=~ m_\phi^2 + {n^2\over R^2}~
\eeq
and where the fermion masses $m_{\psi n}$ and couplings $\hat g_{n,n'}$ are matrices in spinor space,
each with a ``vector'' or ``Dirac'' part (proportional to the identity in spinor space)
and an ``axial'' part (proportional to $\gamma^5$):
\beqn
        m_{\psi n} &=&  m_{\psi n}^{(D)} - i m_{\psi n}^{(A)} \gamma^5\nonumber\\
    \hat{g}_{n,n'} &=&  g_{n,n'}^{(D)} + i g_{n,n'}^{(A)}\,\gamma^5~
\label{massform}
\eeqn
with tree-level values given by
\beq
     \cases{
       m_{\psi n}^{(D)} = m_\psi~,~~~~ m_{\psi n}^{(A)}= n/R & ~\cr
       g_{n,n'}^{(D)} = G/\sqrt{2\pi R}\equiv g~,~~~
      g_{n,n'}^{(A)}= 0~.&~\cr}
\label{treevalues}
\eeq
Note that the axial part of the boson/fermion coupling 
vanishes at tree level.

The situation is nearly identical for a complex scalar field.
Following the same Kaluza-Klein reduction results in a
four-dimensional action  of the form in Eq.~(\ref{genform2}) except
that we no longer 
identify 
$\phi^\ast_n$ with
$\phi_{-n}$,
and we replace $\phi_{r'-r}\to \phi_{r'-r} + \phi^\ast_{r-r'}$ in the
final coupling term.

It may seem, at first glance, that the appearance of the ``axial'' 
$\gamma^5$-terms in the four-dimensional
action violates four-dimensional parity symmetry.
However, it turns out that all terms which are proportional to $\gamma_5$ will also be
odd with respect to $n\to -n$.  As a result, parity will actually be conserved at all
energy scales. 
This, of course, is ultimately a reflection of underyling five-dimensional symmetries.
Indeed, while $\gamma_5$ is odd under the four-dimensional P and CP symmetries,
the quantity $n$ is actually proportional to the momentum component along the fifth dimension.
Thus the quantity $n$ is ``odd'' under P, thereby making the product $\gamma_5 n$ even, as required.

We see, then, that there are five quantities in KK-reduced 5D
Yukawa theory
which are capable of receiving radiative corrections:
$m_{\phi n}$,  
$m_{\psi n}^{(D)}$,  
$m_{\psi n}^{(A)}$,  
$g_{n,n'}^{(D)}$,
and
$g_{n,n'}^{(A)}$.
We shall now explore the one-loop corrections to each of these in turn.

\subsection{Boson KK mass corrections 
\label{bose_shift}}

Regardless of whether the 5D scalar is real or complex, we shall parametrize
the one-loop corrections to the KK boson masses $m_{\phi n}$ in the form 
\beq
          m_{\phi n}^2 ~=~ m_{\phi 0}^2 + \frac{n^2}{R^2} +  \frac{g^2}{4\pi R^2}\, X_{n}^{(m_{\phi}^2)}~
\label{Delt_mph}
\eeq
where $g$ is the universal tree-level coupling in Eq.~(\ref{treevalues}),
as appropriate for a calculation of this order.
Using the techniques developed in Refs.~\cite{1a,1b}, we then find that 
the corresponding functions $\alpha_n$ 
are given by
\beqn
\alpha_{n}(r,v,j) & = & \frac{1}{\pi}
         \biggl\lbrace \left[\rho_n^2 + (1 - 2y_n)|n|\rho_n\right]\,\log(\rho_n^2) \nonumber \\
        && - \left[ \rho_n^2 + (1 - 2y_n)|n|\rho_n +
                3\mathcal{M}_{\phi}^{2}(y_n;m_{\phi}^2)R^2\right]\,
        \log(\rho_n^2 + \mathcal{M}_{\phi}^{2}(y_n;m_{\phi}^2) R^2) \biggr\rbrace 
      \nonumber\\
\label{phn}
\eeqn
where $y_n$ and $\rho_n$ are respectively defined in Eqs.~(\ref{ydef}) and
(\ref{rhodef}) and where
\beq
\mathcal{M}_{\phi}^{2}(y;m_\phi^2) ~=~ m_{\psi}^2 + y(y - 1)m_\phi^2~.
\label{mphi}
\eeq

\begin{figure}[t]
\centerline{ \epsfxsize 2.6 truein \epsfbox {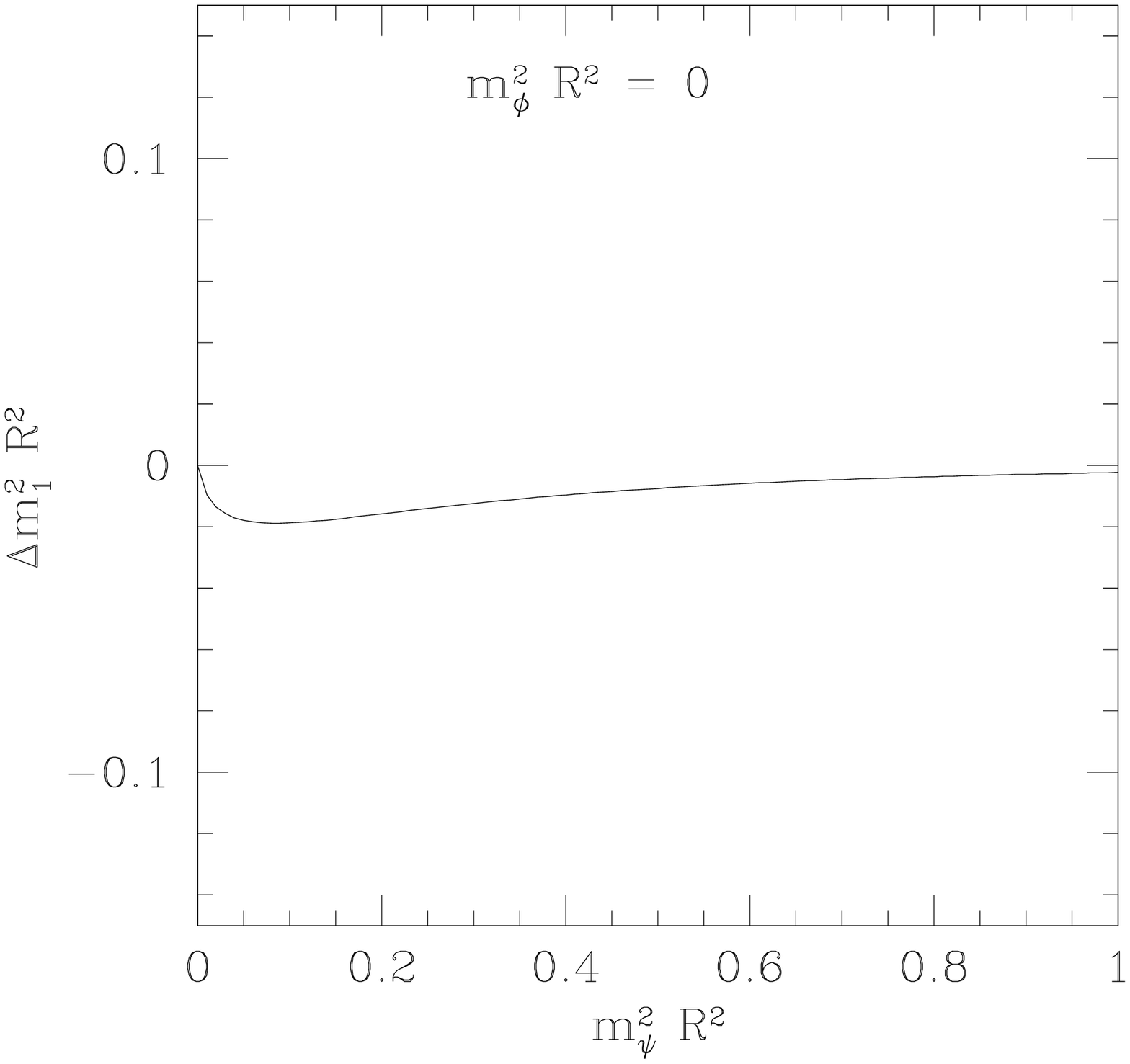} 
             \hskip 0.5 truein \epsfxsize 2.6 truein \epsfbox {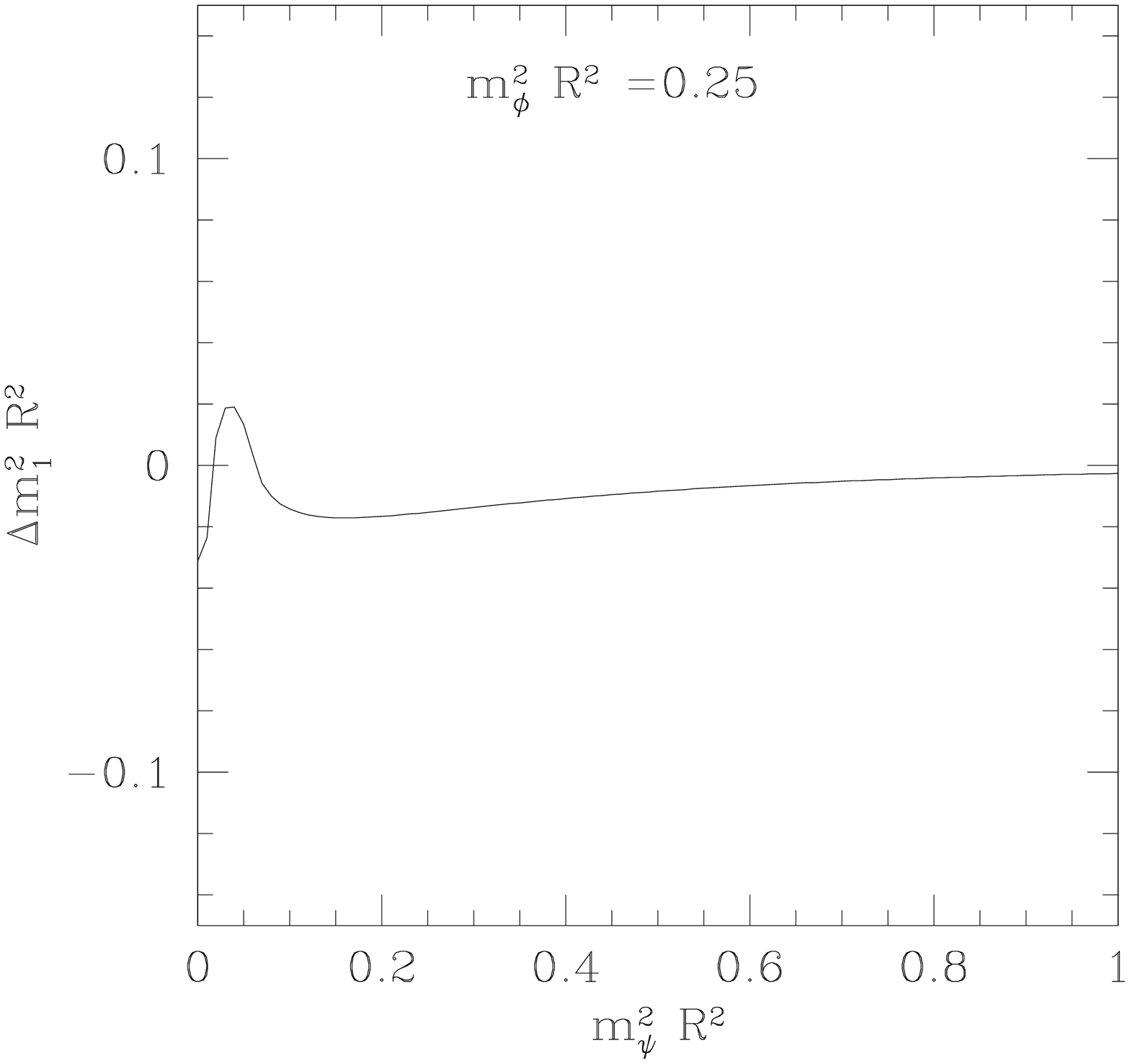} }
\vskip -0.2 truein
\centerline{ \epsfxsize 2.6 truein \epsfbox {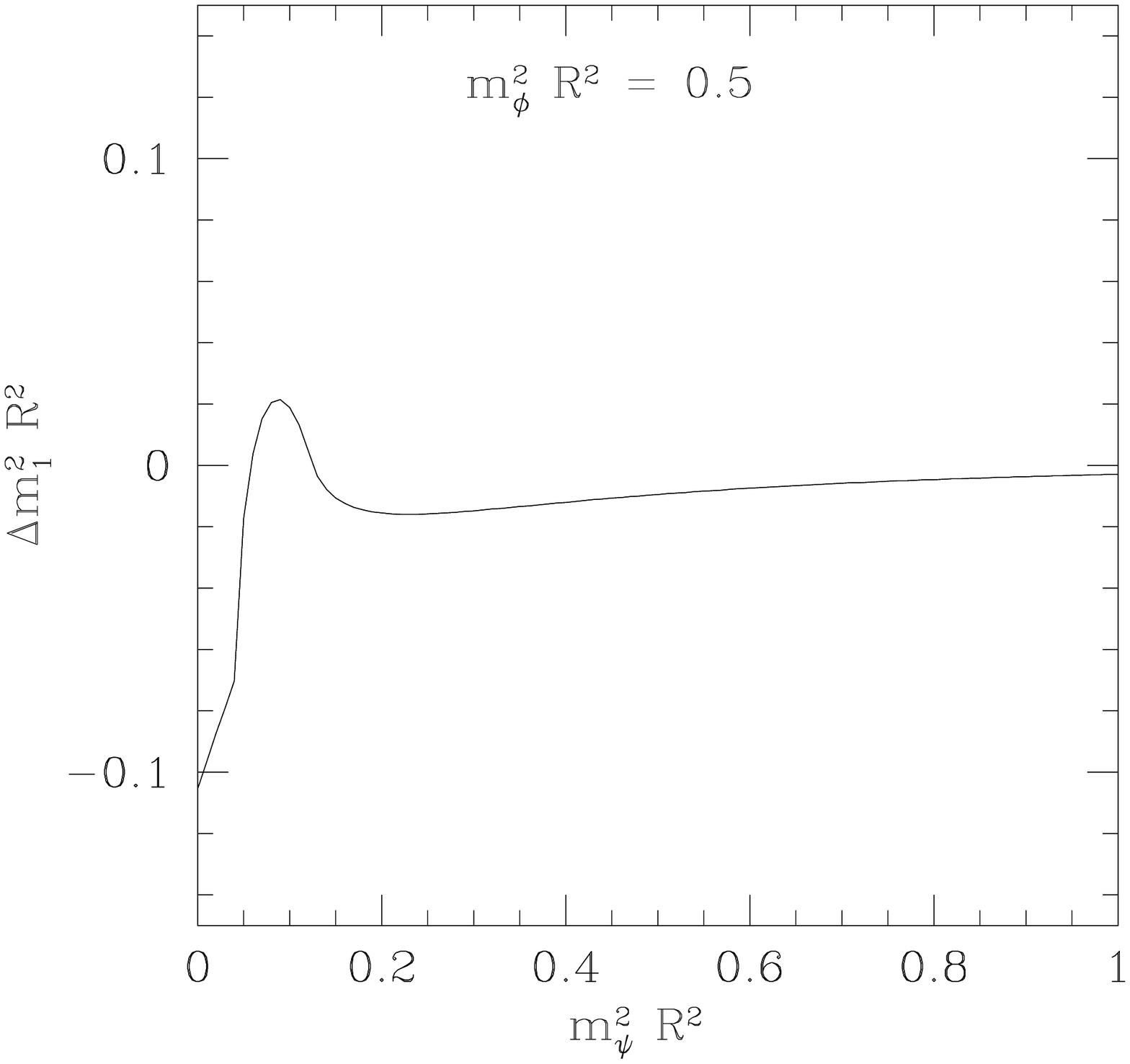} }
\vskip -0.1 truein
\caption{The relative one-loop boson mass corrections $X_1^{(m_\phi^2)}$ between 
  the first-excited KK boson mode and the KK zero mode, 
  plotted as functions of $m_\psi$ for different values of $m_\phi$.}
\label{fig4}
\end{figure}

This compact result contains a wealth of information.
One important feature is the behavior of $X_1^{(m_\phi^2)}$  --- \ie, the radiative
correction
to the mass of the first-excited KK boson relative to the mass of the KK zero mode ---
as a function of the two five-dimensional masses in our problem, $m_\phi$ and $m_\psi$.
This behavior is  
shown in Fig.~\ref{fig4}, where
$X_1^{(m_\phi^2)}$ is plotted as a function of $m_\psi$ for three different
``benchmark'' values of $m_\phi$.
Several features are immediately apparent:
\begin{itemize}
\item  $X_1^{(m_\phi^2)}= 0$ for $m_\phi=m_\psi=0$.
        We shall see, in fact, that this is a general phenomenon for all $X_n^{(m_\phi^2)}$.
\item  $X_1^{(m_\phi^2)}$ is negative when $m_\psi=0$ and $m_\phi\not= 0$.
     This means that the mass splitting between the first-excited KK boson mode 
      and the KK zero mode is {\it reduced}\/ by one-loop radiative corrections --- \ie, 
     these two states begin to approach each other.  Moreover, the magnitude of this effect
      increases with increasing $m_\phi$.  
\item  $X_1^{(m_\phi^2)}\to 0$ as $m_\psi\to \infty$ for all $m_\phi$.
            This occurs  because the functions $\alpha_n$ and $\alpha_0$ in Eq.~(\ref{phn}) 
           approach each other in this limit.  There is therefore no difference in this limit
       between the corrections to the masses of the KK zero mode and first-excited mode --- \ie,
      in this limit the tree-level mass spacing between the zero mode and first excited mode is preserved 
       to one-loop order.
\item  $X_1^{(m_\phi^2)}$ is generally non-monotonic as a function of $m_\psi$.  For $m_\phi$ above
      a critical value,  $X_1^{(m_\phi^2)}$ actually reaches a positive maximum 
      for a value of $m_\psi$ which
      increases with $m_\phi$.
      This non-monotonic behavior emerges as the result of a competition between
      the corrections to the mass of the first-excited KK mode and
      the corrections to the mass of the KK zero mode. 
      Indeed, each of these corrections is individually monotonic.
\item Finally, although it may be somewhat difficult to observe in Fig.~\ref{fig4}, 
      it turns out that $X_1^{(m_\phi^2)}$ 
     actually experiences a {\it kink}\/ (\ie, a slight discontinuous change in slope)
      as a function of $m_\psi$ prior to reaching its maximum value.
      Indeed, this occurs for all $m_\phi >0$.  
     These kinks mark the thresholds
     for the decays of either the KK boson zero mode or the KK first-excited mode. Indeed,
     these thresholds correspond
     to values of $m_\psi$ at which
     the imaginary parts of the diagrams which
     renormalize the scalar masses become zero.
\end{itemize}

\begin{figure}[htb]
\centerline{ \epsfxsize 2.6 truein \epsfbox {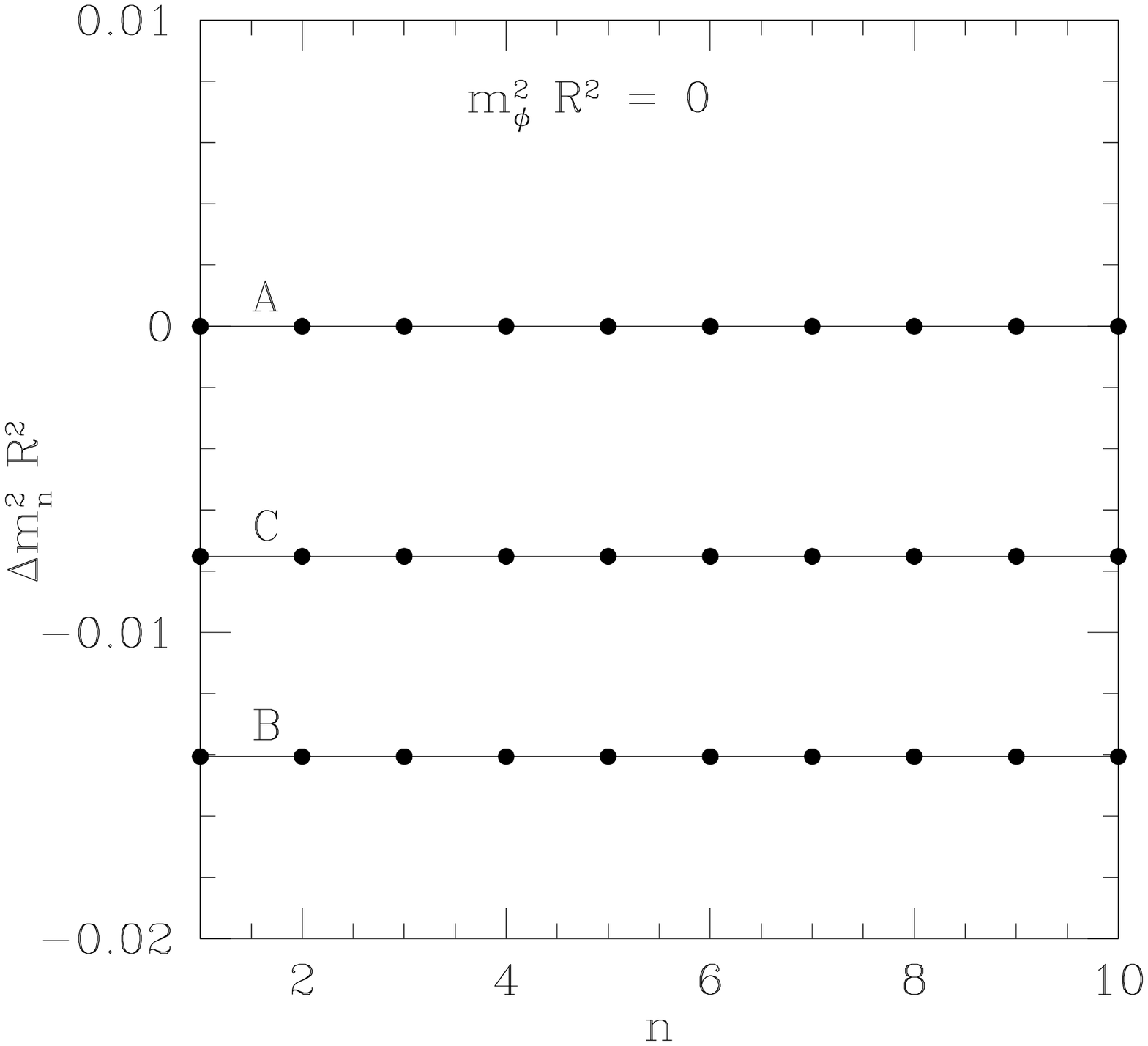} 
             \hskip 0.5 truein \epsfxsize 2.6 truein \epsfbox {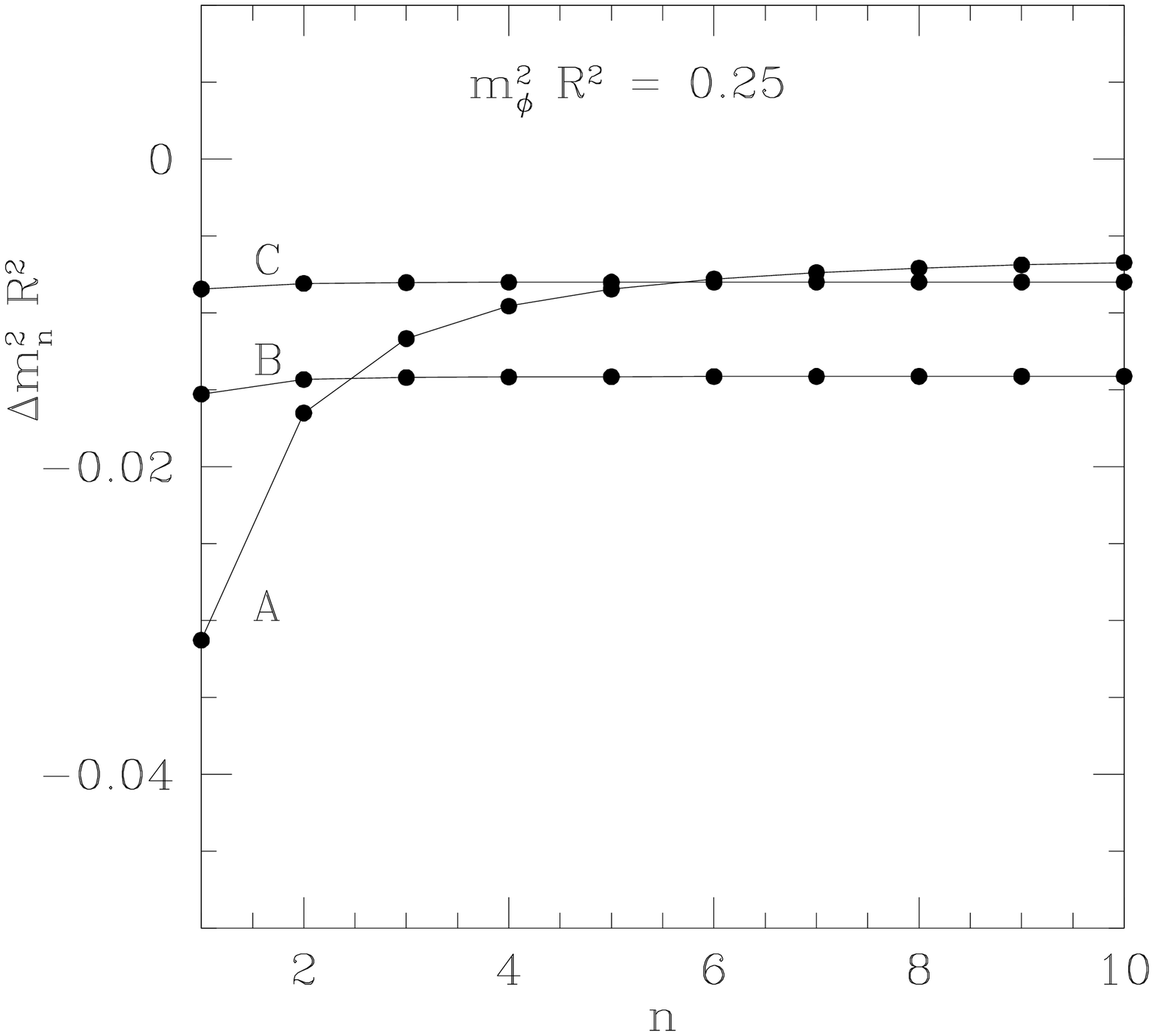} }
\vskip -0.2 truein
\centerline{ \epsfxsize 2.6 truein \epsfbox {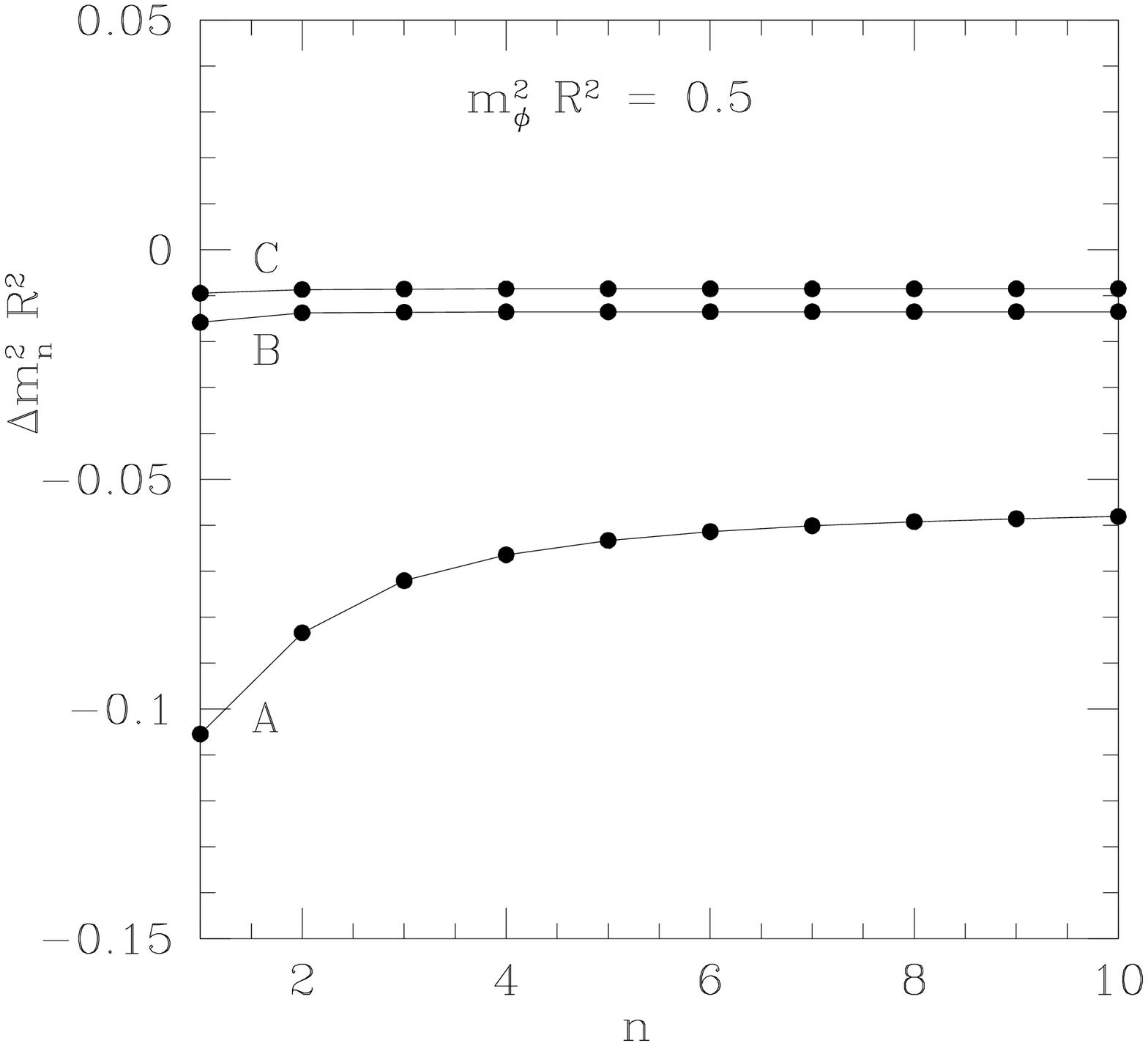} }
\vskip -0.1 truein
\caption{The relative one-loop corrections $X_n^{(m_\phi^2)}$ 
for the KK boson squared masses,
plotted as functions of $n$ for different values of $m_\phi$.
 In each plot, Curves~A, B, and C represent the cases with $m_{\psi}^2 R^2 = 
       \lbrace 0, 0.25, 0.5\rbrace$ respectively --- a naming scheme that will continue 
      to hold for all remaining figures in this paper.}
\label{fig3b}
\end{figure}

We can also examine $X_n^{(m_\phi^2)}$ as functions of $n$.
This behavior is shown in Fig.~\ref{fig3b} for different values of $m_\phi$ and $m_\psi$.
Once again, certain features are readily apparent:
\begin{itemize}
\item For $m_\phi=m_\psi=0$, we find that $X_n^{(m_\phi^2)}=0$ for all $n\geq 0$.
      This is therefore an example of Case~\#1 from the Introduction:  the tree-level form
      of the KK mass relation for the bosonic fields is preserved at one loop.  
      We thus see that it is only the presence of a non-zero five-dimensional mass, 
      either $m_\phi$ or $m_\psi$, which breaks the apparent 5D Lorentz invariance
      as far as the tree-level bosonic spectrum is concerned.
\item For $m_\phi=0$ and $m_\psi\not= 0$, we find that $X_n^{(m_\phi^2)}=$~constant as
      a function of $n$. 
      In fact, this constant depends on $m_\psi$ in a non-monotonic way, hitting
      zero only for $m_\psi=0$ (as discussed in the previous case).  
      This is therefore an example of Case~\#3a from the Introduction:
      all excited KK modes have masses which shift uniformly relative to that of the KK zero mode.
      Thus, all KK modes continue to obey the tree-level mass relation except for the zero mode.
\item  For $m_\phi\not=0$, we find that $X_n^{(m_\phi^2)}$ increases with $n$ but quicky 
         reaches a non-zero asymptote as $n\to \infty$.
     This is therefore an example of Case~\#3b, but with 
        a behavior resembling that of Case~\#3a for the uppermost portions of the KK tower.
\end{itemize}

It should come as no surprise that the radiative corrections to KK masses 
are generically of the form given in Case~\#3a when $n \to \infty$  --- \ie,
that they become independent of $n$ as $n\to\infty$.
The limit of large KK mode numbers corresponds to high momentum
components along the extra dimension, and the discretization of
momentum that arises due to compactification becomes negligible in this limit. 
We therefore expect that the limit of high KK mode numbers should
correspond to an uncompactified theory in which the tree-level KK dispersion
relation holds (signifying the restoration of a full 5D Lorentz invariance).
By contrast, the lower portions of the KK tower
are more sensitive to the discretization of the momentum in the 
compactified dimension.
Thus the 
upper portions of the KK tower have approximately equal mass-squared spacings 
relative to each other, but this pattern does not hold all the way down to
the zero mode.
  
It is also instructive to see how this 
asymptotic behavior of equal spacings emerges analytically.
Towards this end, we can use the $\alpha_n$-functions in Eq.~(\ref{phn})
in order to calculate the contribution to $X_n^{(m_\phi^2)}$ from 
states with a fixed mode number $n'$.  For simplicitly, we shall assume that $n'\gg n$, and
likewise we shall assume that
$n'$ is chosen sufficiently large that $\Lambda \gg m_\phi,m_\psi$ 
where $\Lambda\equiv n'/R$.
We can then expand this contribution in powers of $m/\Lambda$ where $m$
denotes either $m_\phi$ or $m_\psi$, and 
we find that this expansion takes the form
\beqn
&& \frac{g^2}{4\pi^2}
\Biggl\lbrace \left(\frac{7m_{\psi}^2 - m_{\phi}^2}{3 }\right) {1\over \Lambda^2} \nonumber\\
&& ~~~+ \Biggl[ \frac{9m_{\psi}^2 - m_{\phi}^2}{10 R^2} 
 - \frac{11m_{\psi}^4}{2} + \frac{1}{6}\left( -1 + \frac{1}{10n^2}
\right) m_{\phi}^4 + \frac{1}{3}\left( 5 - \frac{1}{5n^2} \right)
m_{\psi}^2 m_{\phi}^2 \Bigg] \frac{1}{\Lambda^4} \nonumber\\
&& ~~~+ \mathcal{O}(m^6 /\Lambda^6 ) \Biggr\rbrace~.
\label{expand2}
\eeqn
Although we have made no assumptions about the size of $n$ itself,
we see that each coefficient in our expansion depends on $n$ only through negative powers.
This is ultimately the source of
the fact that our total mass corrections exhibit a finite, asymptotic limit as $n\to\infty$.
Indeed, although the results in Eq.~(\ref{expand2}) hold only for very large $n'$, 
it turns out that the behavior illustrated in these results is in 
fact completely general, and holds even for smaller values of $n'$
as well.

\subsection{Fermion KK mass corrections 
\label{fermi_shift}}

We now turn to the renormalized masses of the KK fermion modes.
Recall from Eq.~(\ref{treevalues}) that these masses contain both
a vector (or ``Dirac'') component $m_\psi^{(D)}$ and an axial component $m_\psi^{(A)}$.
Parametrizing the one-loop corrections to these masses in the form
\beqn
     m_{\psi n}^{(D)} &=& m_{\psi 0}^{(D)} + \frac{g^2}{4\pi R} X_{n}^{(m_{\psi}^{(D)})} \nonumber\\
     m_{\psi n}^{(A)} &=& \frac{n}{R} + \frac{g^2}{4\pi R}  X_{n}^{(m_{\psi}^{(A)})}~,
\label{linearmass}
\eeqn
we find that the corresponding $\alpha_n$-functions take the forms
\beqn
    X_{n}^{(m_{\psi}^{(D)})}:~~&& 
     \alpha^{(\psi D)}_n ~=~ \frac{m_{\psi}R}{4\pi}(1 + y_n)\log(\rho_n^2 + \mathcal{M}_{\psi}^2 (y_n) R^2)
            \nonumber\\
    X_{n}^{(m_{\psi}^{(A)})}:~~&& 
     \alpha^{(\psi A)}_n ~=~ \frac{{\rm sign}\/(n)}{4\pi}\rho_n 
                   \left[ \log(\rho_n^2 + \mathcal{M}_{\psi}^2 (y_n)R^2) - \log(\rho_n^2)\right]~
\label{fermioncouplings}
\eeqn
where
\beq
  \mathcal{M}_{\psi}^2 (y) ~\equiv~ (y - 1)^2 m_{\psi}^2 + y m_{\phi}^2~.
\label{mpsi}
\eeq
The variables $y_n$ and $\rho_n$ were defined in Eqs.~(\ref{ydef}) and (\ref{rhodef}) respectively.
While these results apply if the five-dimensional scalar $\phi$ is real, 
promoting the 5D scalar to a complex field
merely doubles the values of both of the $\alpha_n$-functions.
Note that the quantity ${\rm sign}\/(n)$ in Eq.~(\ref{fermioncouplings}) is taken to be zero 
when $n = 0$, as a consequence of which the function $\alpha^{(\psi A)}_0$ vanishes.

The results in Eq.~(\ref{fermioncouplings}) describe the
corrections to the masses of the fermion
KK modes.
However, for the sake of comparison with our results for
the boson KK modes,
it will actually be more appropriate to consider the
corresponding corrections to the {\it squared}\/ masses of the fermion
KK modes.
However, given the parametrizations in Eq.~(\ref{linearmass}),
we immediately see that
\beqn
     m_{\psi n}^{(D)2} &=& m_{\psi 0}^{(D)2} + \frac{g^2}{4\pi R^2} Y_{n}^{(m_{\psi}^{(D)})} + {1\over R^2} {\cal O}(g^4) \nonumber\\
     m_{\psi n}^{(A)2} &=& \frac{n^2}{R^2}   + \frac{g^2}{4\pi R^2} Y_{n}^{(m_{\psi}^{(A)})} + {1\over R^2} {\cal O}(g^4)~,
\label{squaredmass}
\eeqn
where the corrections to the squared masses are given to lowest order in $g$ by
\beq
           Y_{n}^{(m_{\psi}^{(D)})} ~=~ 2  m_{\psi 0}^{(D)} R\,     X_{n}^{(m_{\psi}^{(D)})}~,~~~~~
           Y_{n}^{(m_{\psi}^{(A)})} ~=~ 2n \,                       X_{n}^{(m_{\psi}^{(A)})}~.
\label{Ydefs}
\eeq
Indeed, retaining higher orders in $g$ would be incorrect since additional contributions at such orders would
also come from two-loop diagrams, which we have been neglecting.

\begin{figure}[ht]
\centerline{ \epsfxsize 2.6 truein \epsfbox {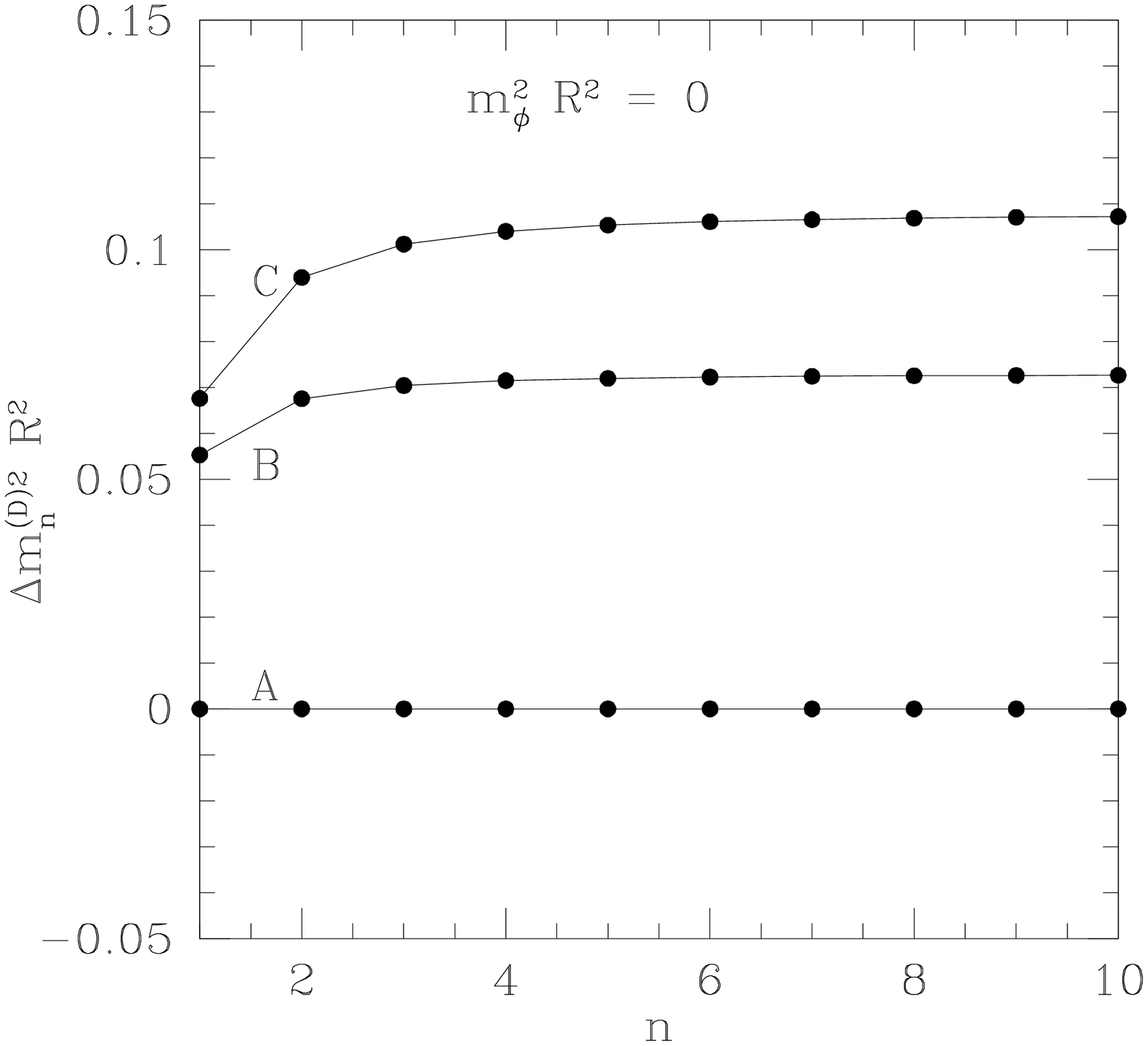} 
             \hskip 0.5 truein \epsfxsize 2.6 truein \epsfbox {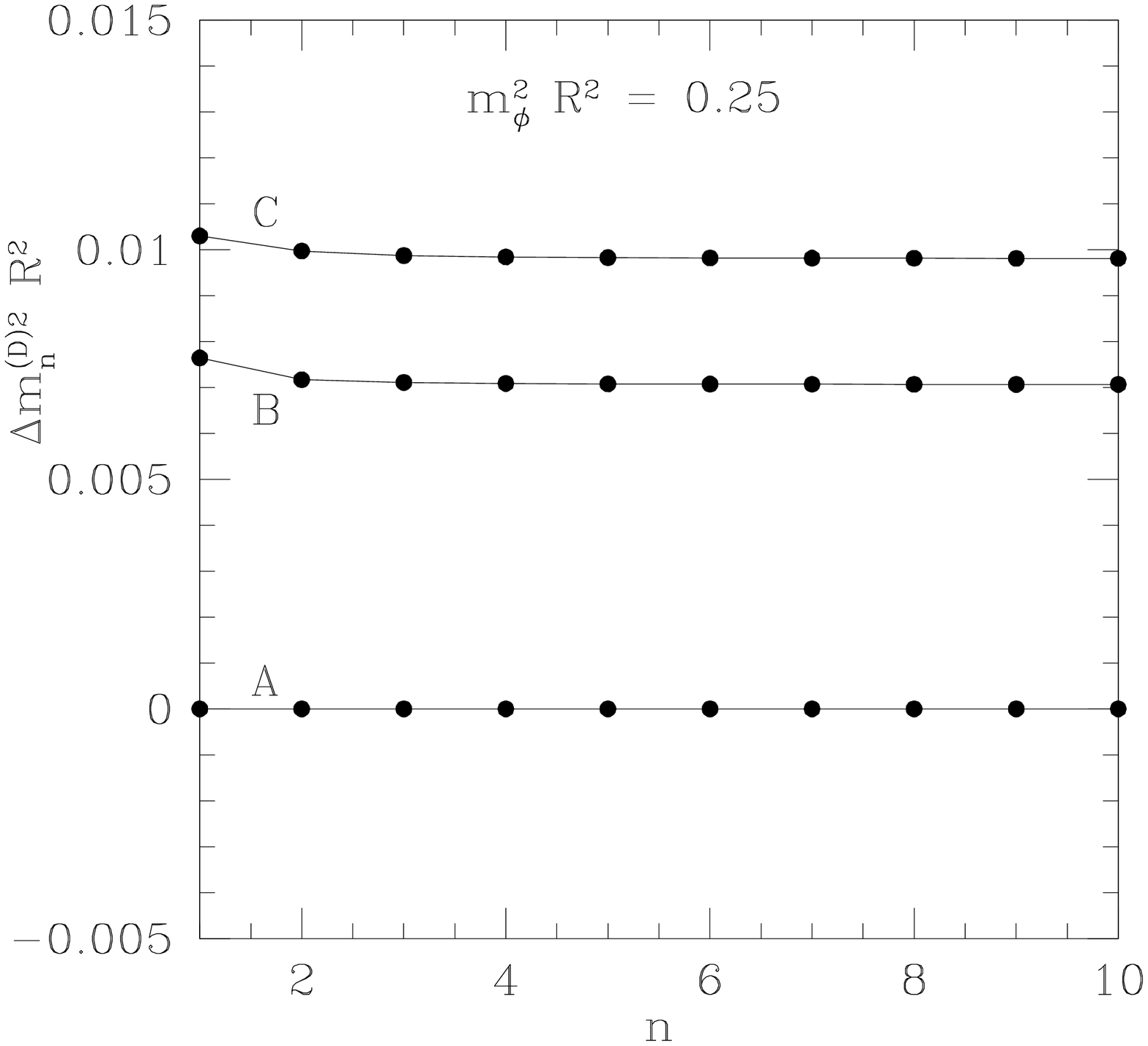} }
\vskip -0.2truein
\centerline{ \epsfxsize 2.6 truein \epsfbox {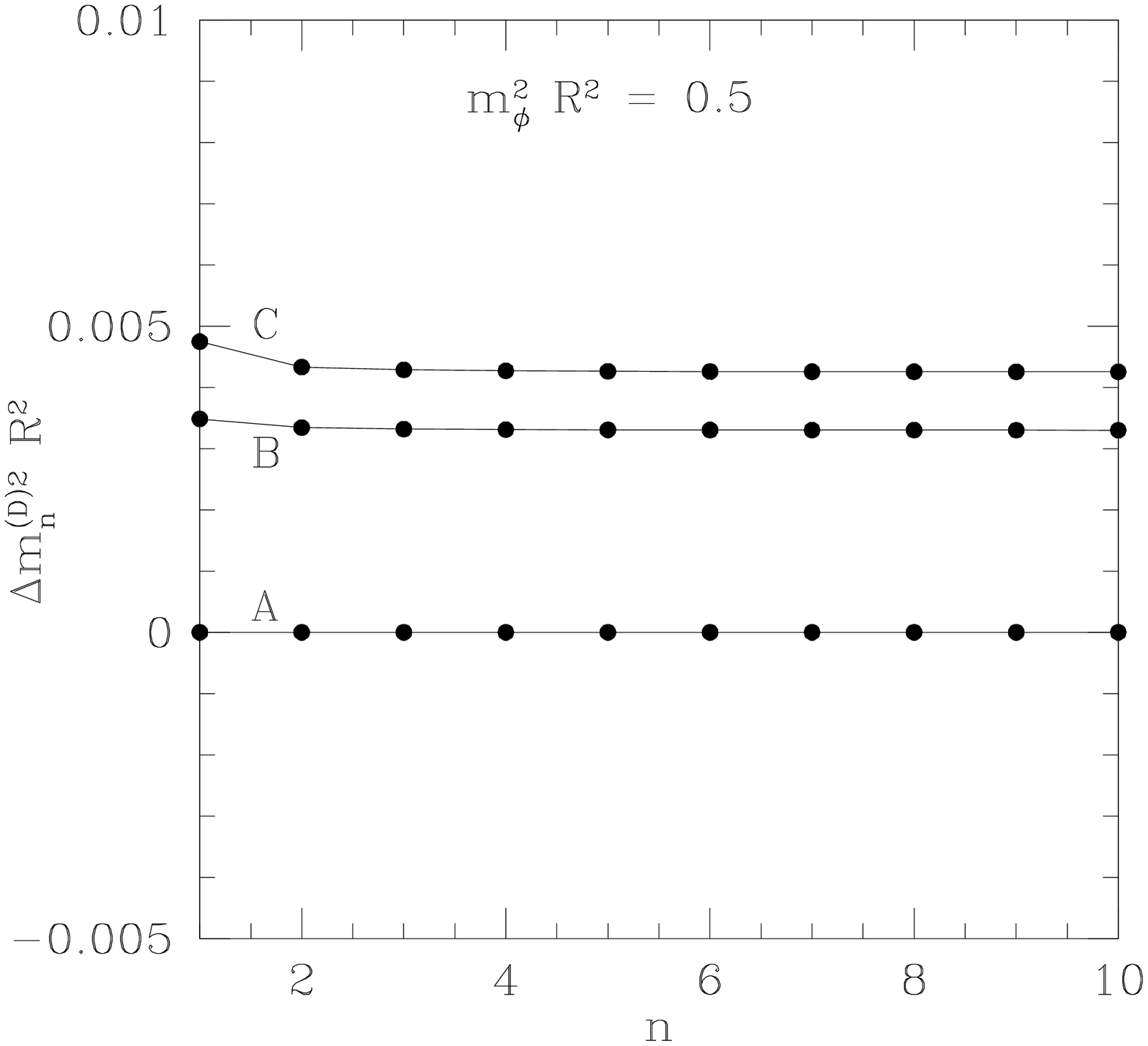} }
\vskip -0.1truein
\caption{The relative one-loop corrections $Y_{n}^{(m_{\psi}^{(D)})}$ 
for the KK fermion squared Dirac masses,
plotted as functions of $n$ for different values of $m_\phi$ and $m_\psi$.}
\label{fig6}
\end{figure}

\begin{figure}[ht]
\centerline{ \epsfxsize 2.6 truein \epsfbox {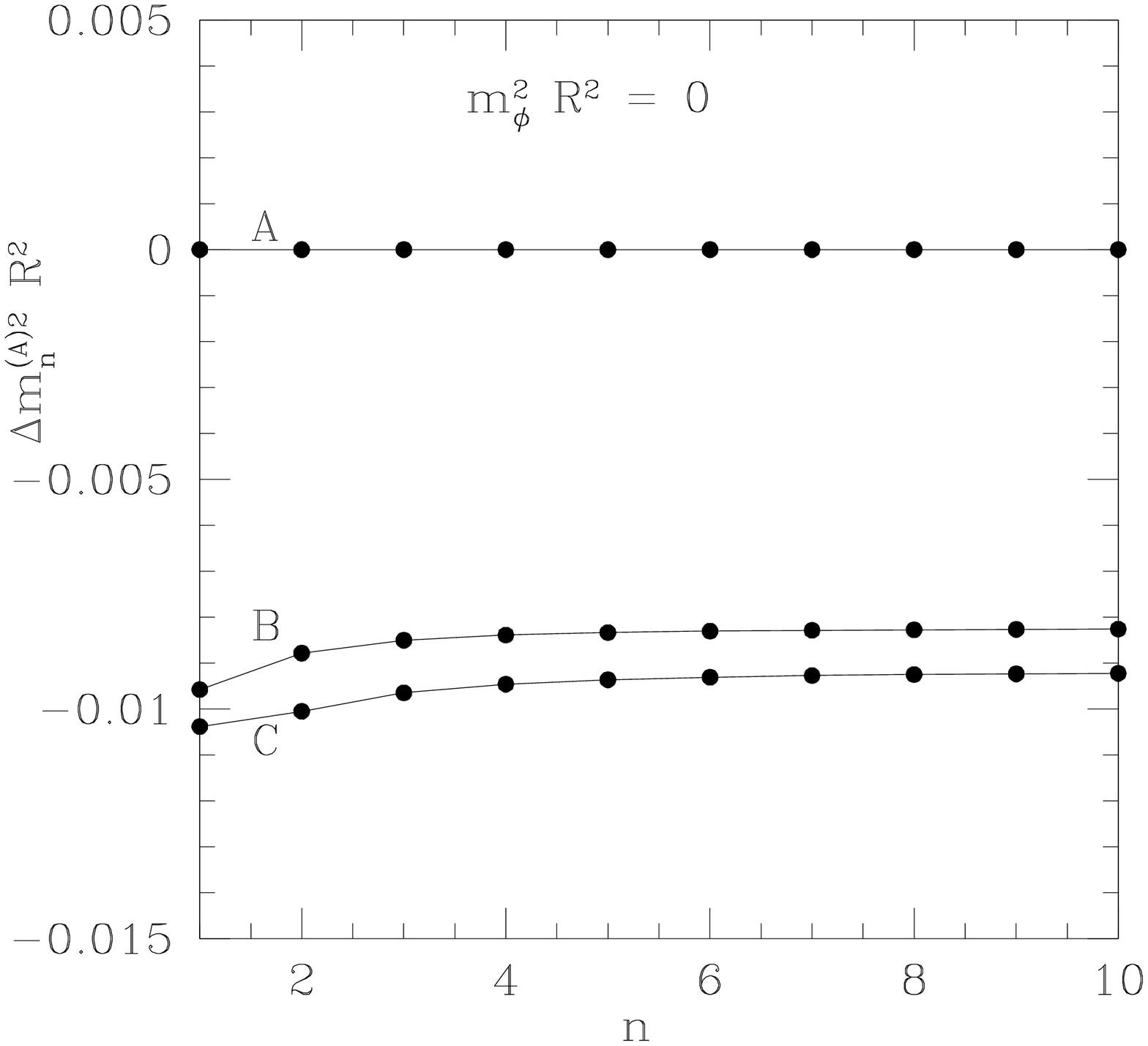} 
             \hskip 0.5 truein \epsfxsize 2.6 truein \epsfbox {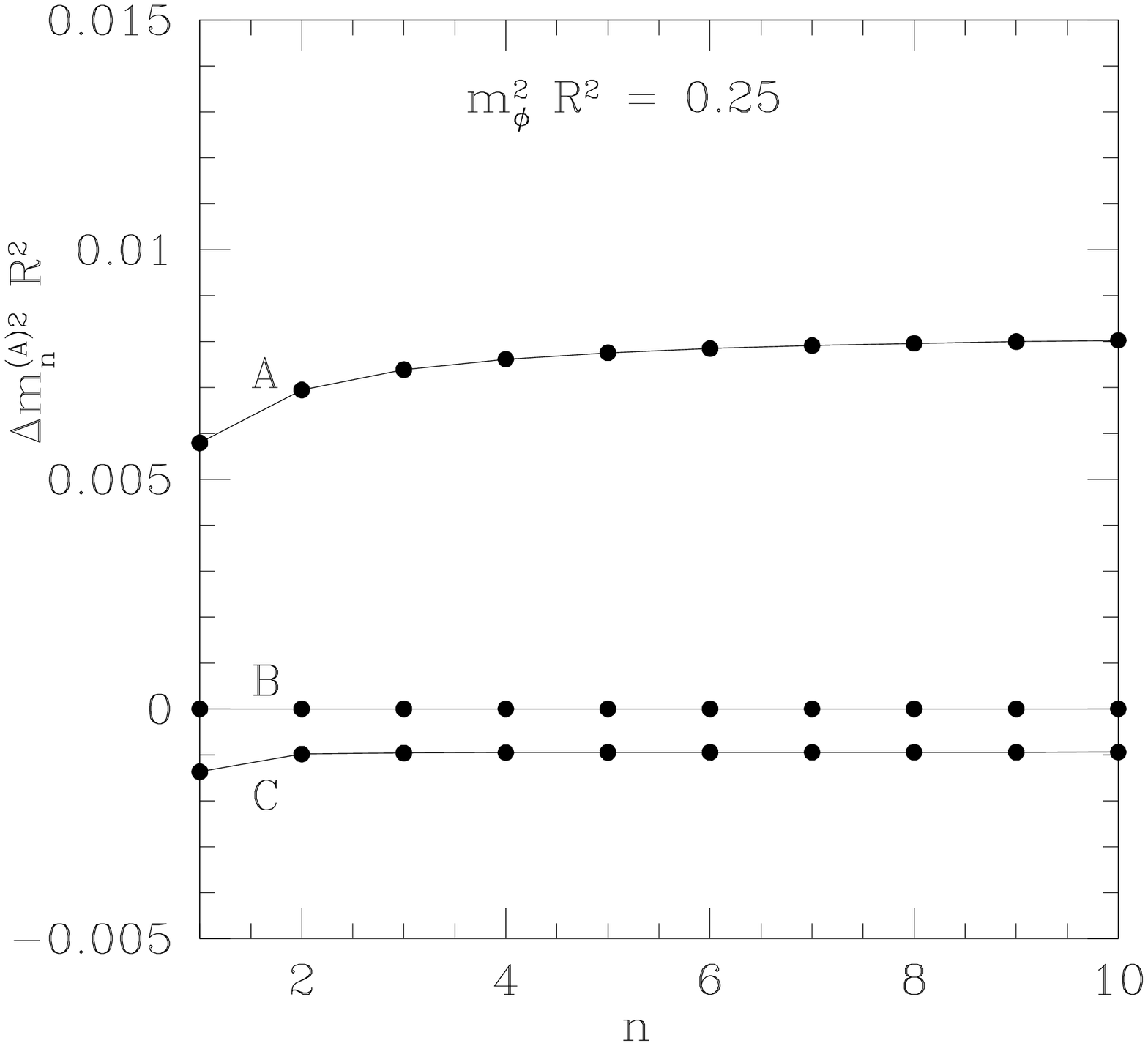} }
\vskip -0.2truein
\centerline{ \epsfxsize 2.6 truein \epsfbox {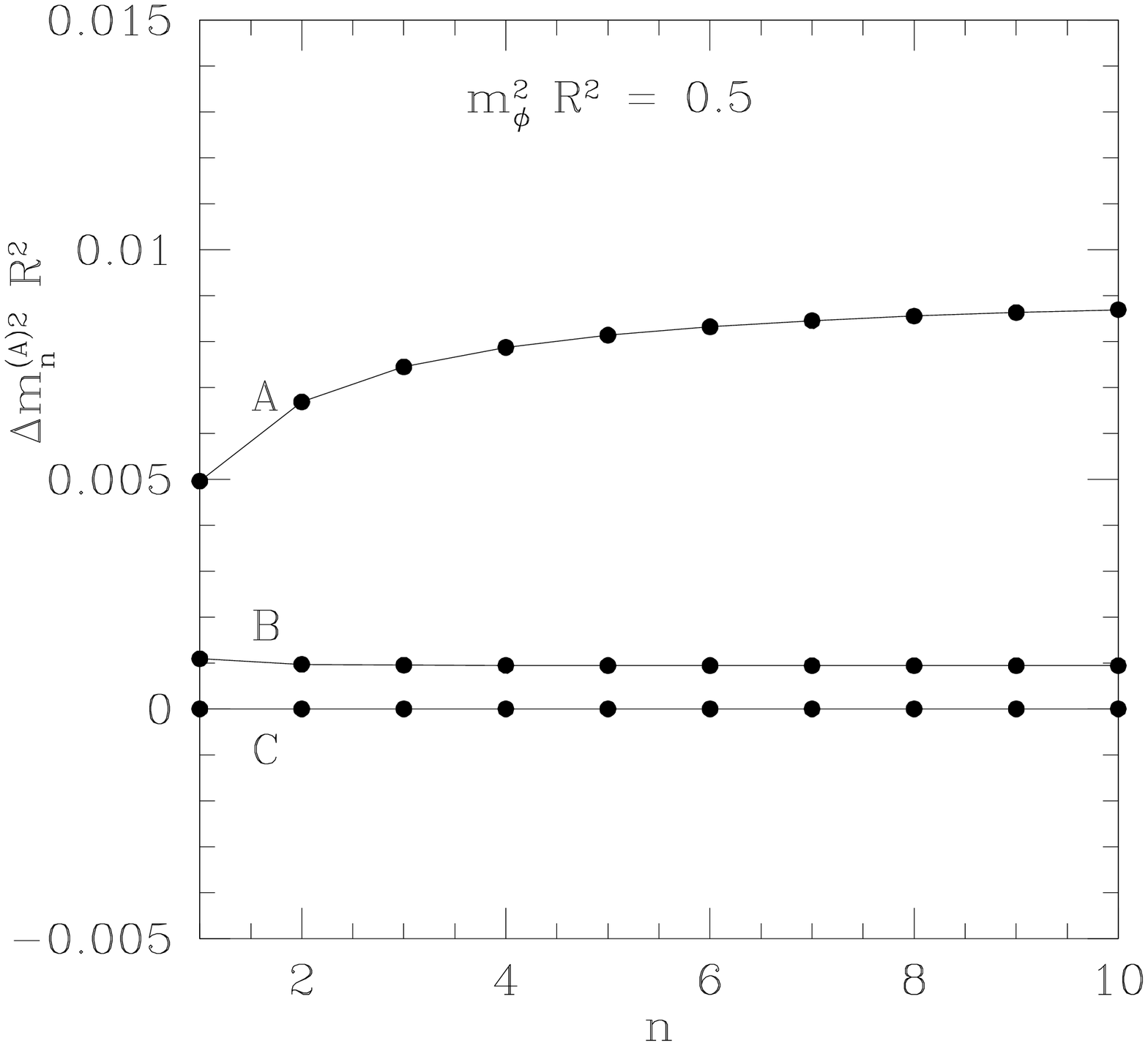} }
\vskip -0.1truein
\caption{The relative one-loop corrections $Y_{n}^{(m_{\psi}^{(A)})}$ 
for the KK fermion squared axial masses,
plotted as functions of $n$ for different values of $m_\phi$ and $m_\psi$.}
\label{fig7}
\end{figure}

\begin{figure}[ht]
\centerline{ \epsfxsize 2.6 truein \epsfbox {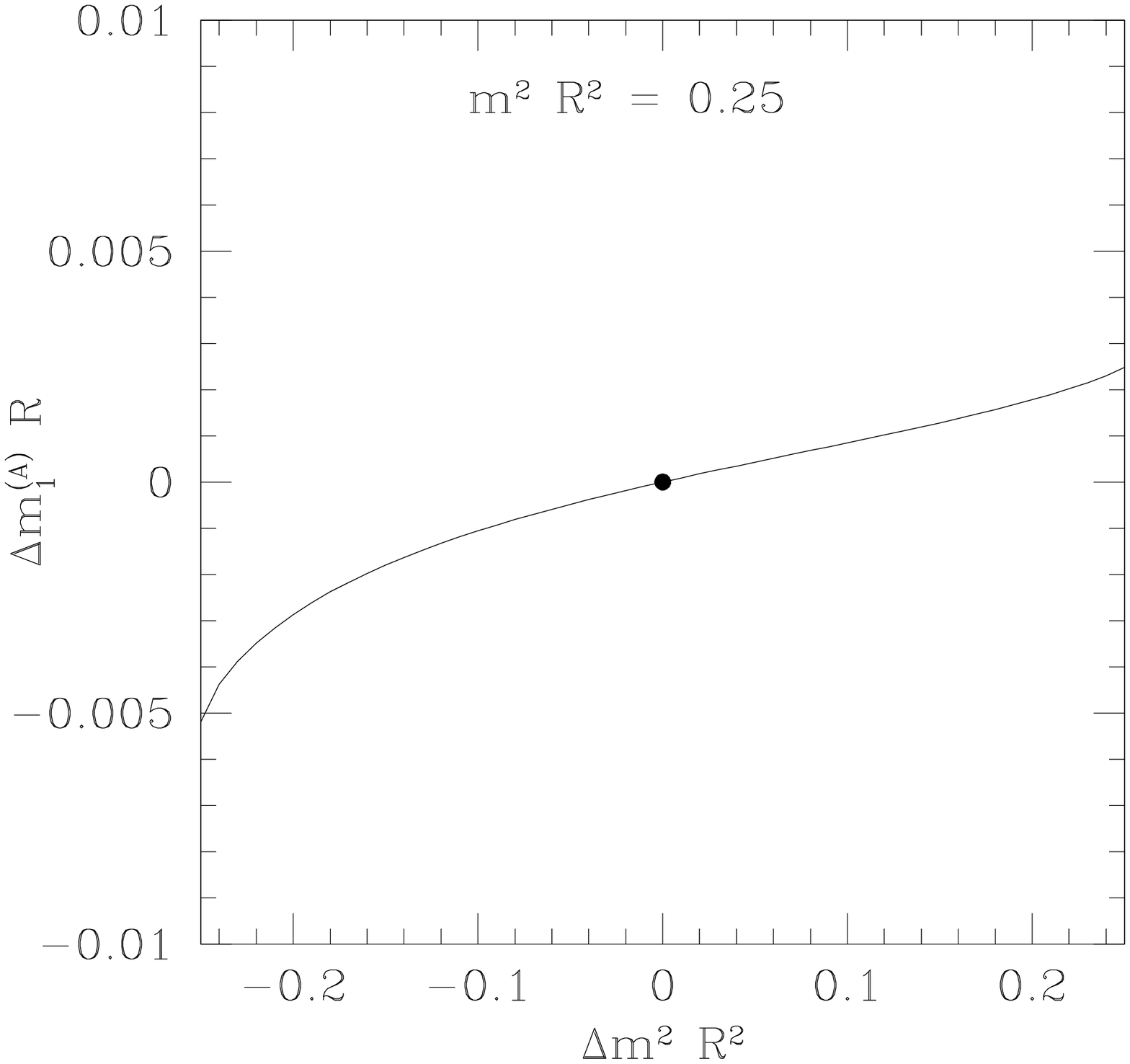} 
        \hskip 0.2 truein
             \epsfxsize 2.6 truein \epsfbox {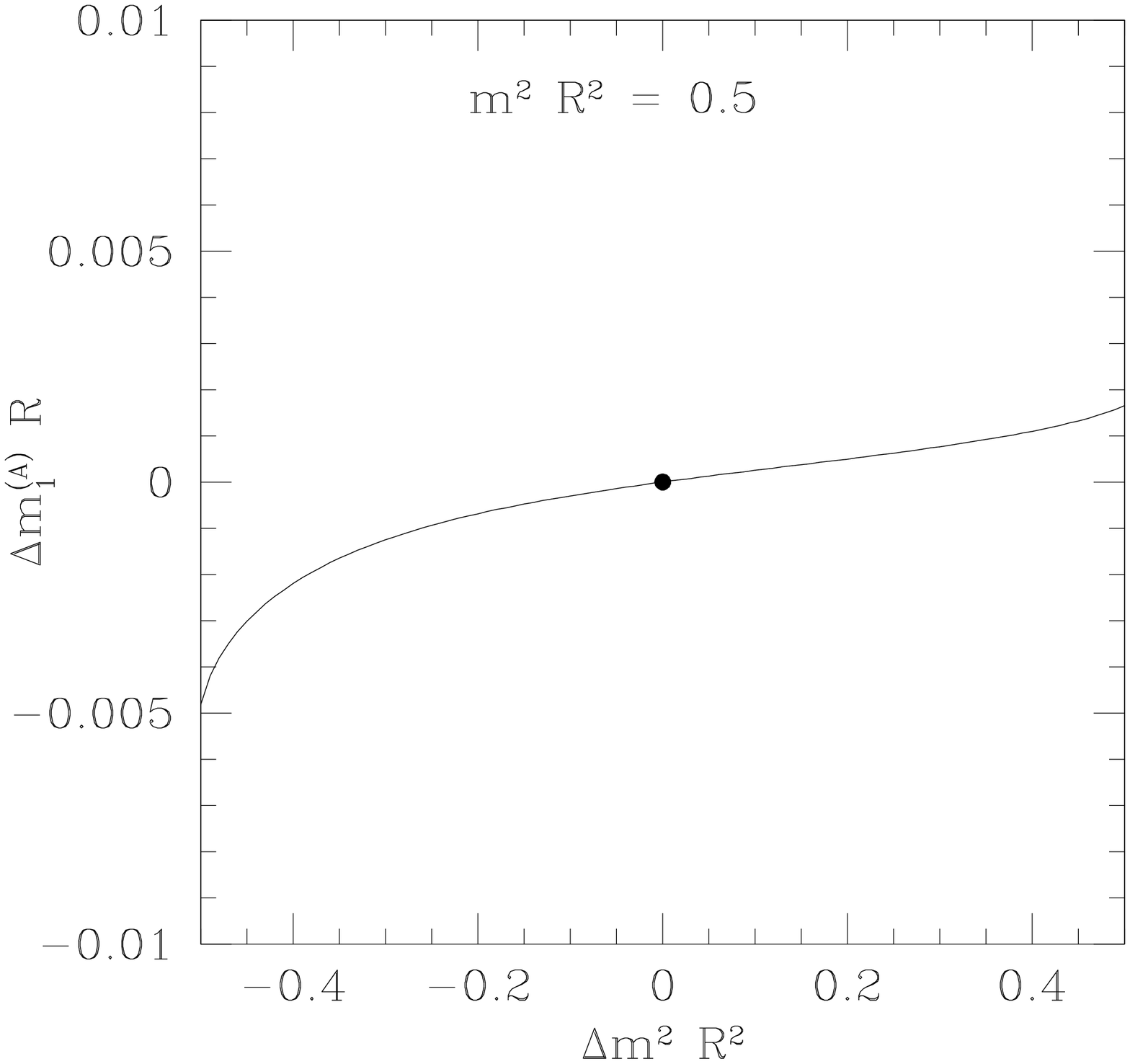} }
\vskip -0.1 truein
\caption{The relative one-loop correction $X_{1}^{(m_{\psi}^{(A)})}$ 
  to the linear axial mass of the first excited KK fermion, plotted versus $\Delta m^2$ 
  for fixed values of $m^2$.
  The quantities $m^2$ and $\Delta m^2$ are defined such that 
  $m_{\psi}^2 = m^2 - \Delta m^2$, and $m_{\phi}^2 = m^2 + \Delta m^2$.  
  As in Fig.~\ref{fig7}, we see that these corrections are 
  positive if $m_\phi>m_\psi$, negative if $m_\phi<m_\psi$, and zero if $m_\phi=m_\psi$.}
\label{fig9}
\end{figure}

\begin{figure}[ht]
\centerline{ \epsfxsize 2.6 truein \epsfbox {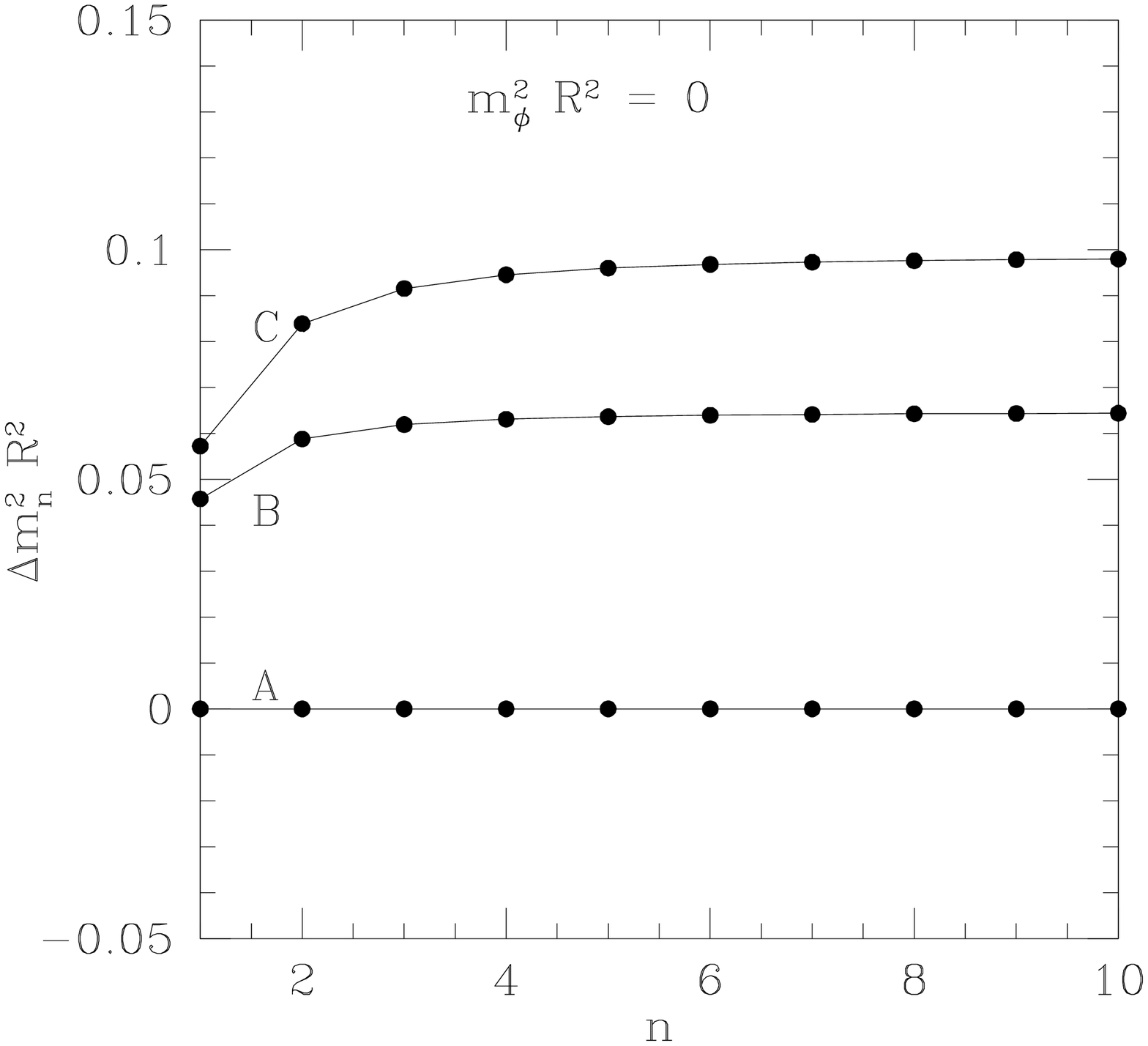} 
             \hskip 0.5 truein \epsfxsize 2.6 truein \epsfbox {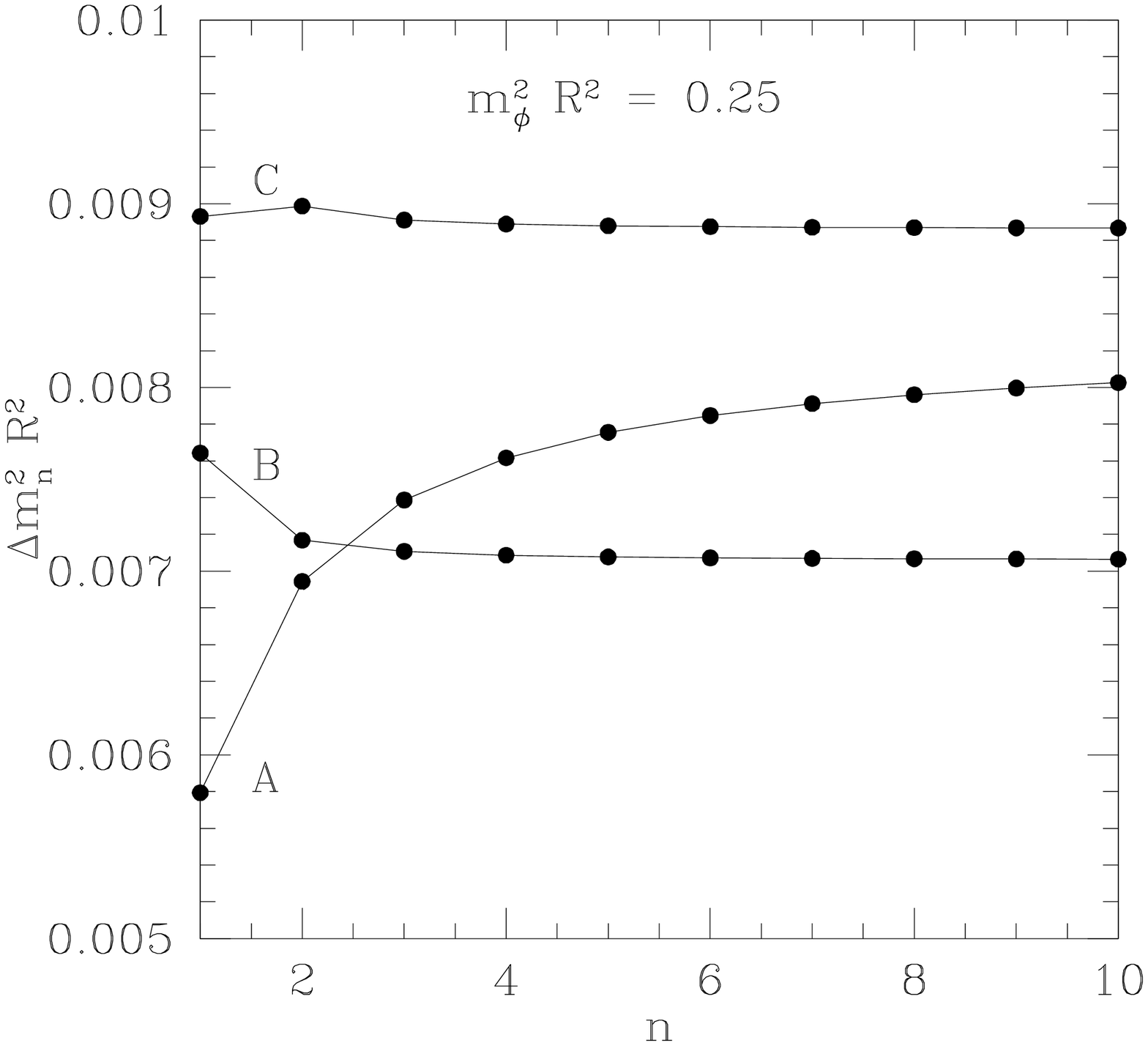} }
\vskip -0.2truein
\centerline{ \epsfxsize 2.6 truein \epsfbox {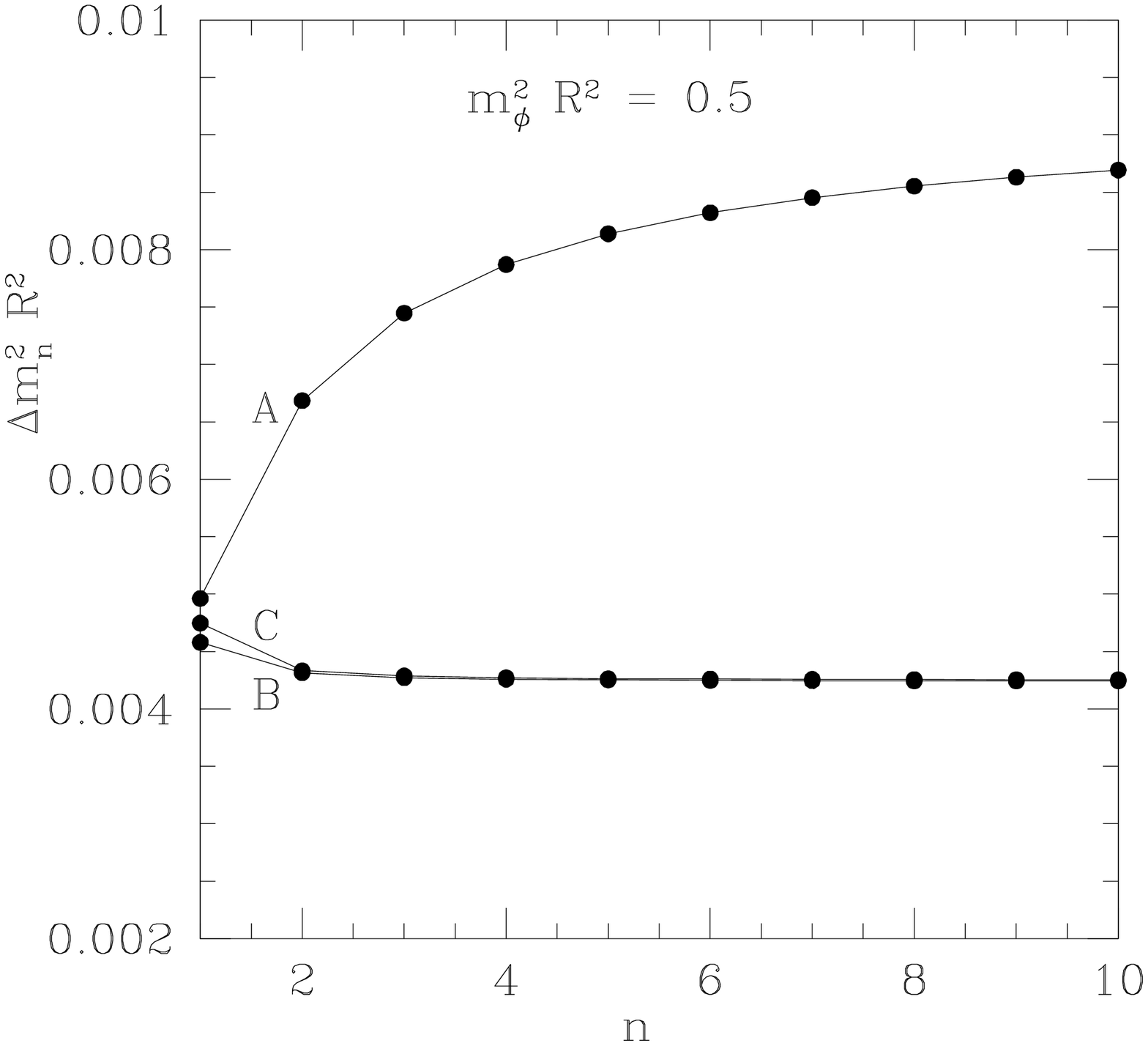} }
\vskip -0.1truein
\caption{The relative one-loop corrections 
$Y_n^{(m_\psi)}$ 
for the physical KK fermion masses,
plotted as functions of $n$ for different values of $m_\phi$ and $m_\psi$.}
\label{fig5}
\end{figure}

The corrections to the squared Dirac masses are shown in
Fig.~\ref{fig6}.  Likewise, corrections to the squared axial 
masses are shown in Fig.~\ref{fig7}. 
As we observe from these figures, the Dirac and axial
mass corrections do exhibit certain common behaviors.
For example, in both cases these corrections are monotonic
with mode number $n$, and they each approach constant values
as $n\to \infty$.  

However, there are also certain crucial differences between
the behaviors of the Dirac and axial mass corrections.
The Dirac corrections, for example, vanish if $m_\psi=0$
(regardless of the value of $m_\phi$);  thus it is the fermion
bare mass $m_\psi$ which is responsible for triggering a non-zero
one-loop mass correction.     
Likewise, the Dirac corrections are positive and increase 
as functions of $m_\psi$, while they generally {\it decrease}\/ as 
functions of $m_\phi$ (with $m_\psi$ held fixed).

By contrast, the axial mass corrections are positive if $m_\phi>m_\psi$,
negative if $m_\psi<m_\phi$, and zero if $m_\psi=m_\phi$.
Indeed, the behavior of the correction $X_1^{(m_\psi^{(A)})}$ to the linear
axial mass of the first-excited KK fermionic state is shown in Fig.~\ref{fig9}
as a function of the difference $m_\phi^2-m_\psi^2$, and we see that
this function is positive when this difference is positive, negative
when this difference is negative, and zero precisely when this difference
is zero.

It is an interesting phenomenon that the axial mass corrections vanish for 
$m_\psi=m_\phi$.
It is straightforward to demonstrate this explicitly at one-loop
order using the expressions for the mass corrections 
given above, and one finds that this results from 
a cancellation between the effects of the different KK boson and fermion
propagators in the loop.
This suggests a possible supersymmetric origin for this cancellation,
and indeed we observe that although the Yukawa theory 
under study here is not supersymmetric, the one-loop corrections to
the fermion masses in this Yukawa theory are equivalent (up to an overall multiplicative constant)
to the corresponding corrections in a supersymmetric Yukawa theory, provided $m_\psi=m_\phi$.
This is significant because supersymmetry forbids KK fermions from accruing
axial mass corrections.

Finally, we observe
that 
the corrections to the
axial fermion masses $m_{\psi n}^{(A)}$ are also odd functions of the mode number $n$.
Although this is not evident from the plots in Fig.~\ref{fig7},
this result follows directly as the consequence of the 
analytic expression for the axial mass correction given in Eq.~(\ref{fermioncouplings}):
the prefactor ${\rm sign}(n)$ is odd under $n\to -n$, while the rest of the expression
is manifestly even under $n\to -n$.
This property is a direct consequence of 
the overall P and CP symmetries of our original five-dimensional theory. 
As a corollary, this symmetry protects the fermion zero
mode from gaining an axial mass.

Thus far, we have discussed the corrections
to the individual Dirac and axial components of the 
KK fermion masses.
However, for many purposes the important quantities are actually the 
total physical fermionic masses themselves ---
\ie, the masses corresponding to the poles in the KK fermion propagators.
In general, the squares of these masses are the 
sums of the squares of the two individual mass components:
\beq
       m_{\psi n}^2 ~\equiv~ m_{n\psi}^\dagger m_{n\psi} ~=~
    m_{\psi n}^{(D)2} + m_{\psi n}^{(A)2}~,
\label{fermi_mass}
\eeq
where $m_{\psi n}$ is the fermion mass given in Eq.~(\ref{massform}). 
It then follows from Eq.~(\ref{squaredmass}) 
that the corrections to this mass take the form
\beq
     m_{\psi n}^2  ~=~  m_{\psi 0}^{2} + \frac{n^2}{R^2}   
             + \frac{g^2}{4\pi R^2}  Y_n^{(m_\psi)} + {1\over R^2} {\cal O}(g^4)~
\label{totm}
\eeq
where we have recognized $m_{\psi 0}^2= m_{\psi 0}^{(D)2} + {\cal O}(g^4)/R^2$ 
and where
\beq
      Y_n^{(m_\psi)}~\equiv~ Y_{n}^{(m_{\psi}^{(D)})} + Y_{n}^{(m_{\psi}^{(A)})}~.
\label{totY}
\eeq
These corrections are shown in Fig.~\ref{fig5}.

Unlike the individual corrections to the Dirac and axial
mass components, these overall corrections do not behave as simple monotonic functions
of the bare masses $m_\phi$ and $m_\psi$.  
This non-trivial behavior ultimately 
arises as the result of 
a competition between the contributions from
the Dirac and axial corrections in Eq.~(\ref{fermi_mass}). Indeed, as evident
in Figs.~\ref{fig6} and~\ref{fig7}, these corrections to the squared
Dirac and axial masses vary in opposite directions with respect to the
fermion bare mass.
We also observe that 
these corrections are also generally largest when $m_\phi=0$. 
This enhancement arises due to the fact that the logarithms in the Dirac and axial corrections become
large when their arguments tend to zero. 
We nevertheless see that these corrections all approach constant values as $n\to \infty$,
indicating that the 
the uppermost portions of the KK tower effectively behave according to
Case~\#3a from the Introduction.
We also observe that these corrections vanish only when $m_\phi = m_\psi = 0$.
This is then an example of Case~\#1.

\subsection{Yukawa coupling corrections
\label{coup_shift}}

Finally, we consider the one-loop corrections to the Yukawa
coupling.
Like the coupling in the $\lambda \phi^4$ theory discussed in Sect.~3, we shall express
the Yukawa coupling and its one-loop corrections as functions of 
a canonical (non-Wilsonian) renormalization scale $\mu$, which we shall here take
to be the squared five-momentum of the scalar mode (\ie, $\mu^2 = -Q^2$ where $Q^M$ is the
scalar five-momentum).
This in some sense defines the energy
of the experiment through which 
this coupling is measured.
However, unlike the case of $\lambda \phi^4$ theory,
the results for the one-loop coupling corrections here are more
complicated due to several factors, including the presence of non-zero   
field-strength renormalizations and the existence of relevant Feynman
diagrams involving more than a single Feynman parameter.  Neither of these
features appeared in the $\lambda \phi^4$ theory at one-loop order.
Moreover, as indicated in Eq.~(\ref{massform}), the Yukawa coupling  
actually has two independent components, one ``vector'' (or Dirac) and the other axial.

Despite these complications, we can parametrize the one-loop corrections to these Yukawa
coupling components in the form
\beqn 
  g_{n_1,n_2}^{(D)} &=&  g_{00}^{(D)} + {g^3\over 8\pi^{3/2}}\left[
             L^{(D)}_{n_1,n_2} +  Z^{(\phi)}_{n_2-n_1} + Z^{(\psi)}_{n_1} + Z^{(\psi)}_{n_2} \right]\nonumber\\
  g_{n_1,n_2}^{(A)} &=& {g^3\over 8\pi^{3/2}}\, L^{(A)}_{n_1,n_2}~, 
\label{coupforms}
\eeqn
where $g$
is defined in Eq.~(\ref{treevalues}).
In Eq.~(\ref{coupforms}),
the quantities $Z^{(\phi,\psi)}$ represent the contributions from bosonic 
and fermionic field-strength renormalizations, 
while $L^{(D,A)}$ represent   
those parts of the appropriate one-loop vertex renormalization 
diagram which are proportional to ${\bf 1}$ and $\gamma^5$ respectively in spinor space. 
Note, in particular, that what we are denoting $Z^{(\phi,\psi)}$ are merely {\it contributions}\/ from the
field-strength renormalizations;  they are not the complete renormalizations themselves.
As might be expected, field-strength renormalizations yield corrections to Dirac (vector) couplings 
but not the axial couplings.  In this connection, we observe that there 
were no one-loop field-strength renormalization contributions to the 
analogous coupling corrections in the $\lambda \phi^4$ case because the appropriate
loop integral in the $\lambda\phi^4$ case 
was completely independent of the momentum on the external leg.
This is ultimately the same reason that 
the KK mass relation for the $\phi$ fields in the $\lambda \phi^4$ theory 
was invariant to this order.

Given the parametrization in Eq.~(\ref{coupforms}), our results are as follows.
The field-strength renormalization contributions $Z^{(\phi,\psi)}$ 
take the standard form in Eq.~(\ref{Xform}),
where the corresponding $\alpha_n$-functions are given by
\beqn
 Z_n^{(\phi)}:&~~~&
 \alpha_n ~=~ \frac{1}{\sqrt{\pi}} 
   \, y_n(1 - y_n)\Bigg[ 3\log 
  \left(\rho_n^2 + \mathcal{M}^{2}_{\phi}(y_n;\mu^2)R^2 \right) \nonumber\\
     &&~~~~~~~~~~~~~~~~~~~~~~~+ \frac{(1 - 2y_n)|n|\rho_n +
   2\mathcal{M}^{2}_{\phi}(y_n;\mu^2)R^2}{\rho_n^2 +
    \mathcal{M}^{2}_{\phi}(y_n;\mu^2)R^2}\Bigg]~ \nonumber\\
 Z_n^{(\psi)}:&~~~&
 \alpha_n ~=~ 
 \frac{1}{4\sqrt{\pi}}\, y_n \left[ \log \left( \rho_n^2 + \mathcal{M}^{2}_{\psi}(y_n)R^2 \right) 
          + \frac{2(y_n^2 - 1)m_{\psi}^2 R^2}{\rho_n^2 + \mathcal{M}^{2}_{\psi}(y_n)R^2}\right]~.
\label{phin}
\eeqn
Note that the quantities $y_n$, $\rho_n$, 
$\mathcal{M}^{2}_{\phi}$, and $\mathcal{M}^{2}_{\psi}$
are defined in 
Eqs.~(\ref{ydef}), (\ref{rhodef}),
(\ref{mphi}), and (\ref{mpsi})
respectively.
Also note that on-shell renormalization conditions for $Z^{(\psi)}$ have been applied in obtaining
Eq.~(\ref{phin}). 

The situation is significantly more complex for the contributions 
$L^{(D,A)}$ coming from the
vertex renormalizations because the relevant diagrams 
in this case involve two Feynman parameters rather than just one.
However, it turns out that there does exist a simple closed form for these
corrections which is analogous to that in Eq.~(\ref{Xform}) when 
either $n_1$ or $n_2$ is zero.  For concreteness, let us assume that $n_2$ is zero.
In such cases, Eq.~(\ref{Xform}) is replaced by   
\beq
 L_{n_1,0} ~=~ 
    \sum_{r = -\infty}^{\infty}
    \frac{1}{|n_1|}\sum_{j = 0}^{|n_1| - 1} 
    \int_{0}^{1}dv_1 \int_{0}^{1}dv_2 \, \Big[
    \alpha_{n_1,0}(r,v_1,v_2,j) -
    \alpha_{0,0}(r,v_1,v_2) \Big]~. 
\label{Ln0}
\eeq
We then find that the corresponding $\alpha_{n_1,0}$-functions are given by
\beqn
 L_{n_1,0}^{(D)}:&~~~&
 \alpha_{n_1,0} ~=~ 
         \frac{1}{\sqrt{\pi}}(1 - y_{n_1})
         \Bigg\lbrack \log(\rho_{n_1}^2 + \mathcal{M}^{2}_{g}(y_{n_1},y_0';\mu^2)R^2) \nonumber \\
&& ~~~~~~~~~~~~~~~~~~~~+ \frac{1}{2} \, \frac{\mathcal{M}_{g}^2(y_{n_1},y_0';\mu^2)R^2 
           + (2 - v_1 - v_2)^2 m_{\psi}^2 R^2}{\rho_{n_1}^2 +
  \mathcal{M}_{g}^2 (y_{n_1},y'_0;\mu^2)R^2}  \Bigg\rbrack
            \nonumber \\
 L_{n_1,0}^{(A)}:&~~~&
 \alpha_{n_1,0} ~=~ 
 -\frac{1}{\sqrt{\pi}}(1 - y_{n_1})(2 - y_{n_1} - y_{0}')\,  \frac{\rho_{n_1}\, m_\psi R}{\rho_{n_1}^2 + 
           \mathcal{M}_{g}^2 (y_{n_1},y'_0;\mu^2)R^2} ~,\nonumber\\
\label{ag2n}
\eeqn
where $y_n$ and $\rho_n$ are defined as in Eqs.~(\ref{ydef}) and (\ref{rhodef})
except with $v$ replaced by $v_1$, 
where $y_0'\equiv v_2 (1-y_{n_1})$,
and where
\beq
   \mathcal{M}^{2}_{g}(y,y';\mu^2) ~\equiv~ (y + y')^2 m_{\psi}^2 + (1 - y - y')m_{\phi}^2 - y y' \mu^2~.
\label{Mg}
\eeq

The above expressions for the $\alpha$-functions 
assume that $n_1\not=0$ and $n_2=0$. However, analogous results
exist when $n_1 =0$ and $n_2\not=0$.  
Likewise, the results listed above 
apply when the 5D
scalar in our theory is real. When this field is complex, by contrast, the
$\alpha$-functions corresponding to the $Z^{(\psi)}$ and $L^{(D,A)}$ corrections double,
while the 
$\alpha$-function
corresponding to the $Z^{(\phi)}$ correction remains invariant.

\begin{figure}
 ~\vskip -0.2 truein
\centerline{
   \epsfxsize 2.5 truein \epsfbox {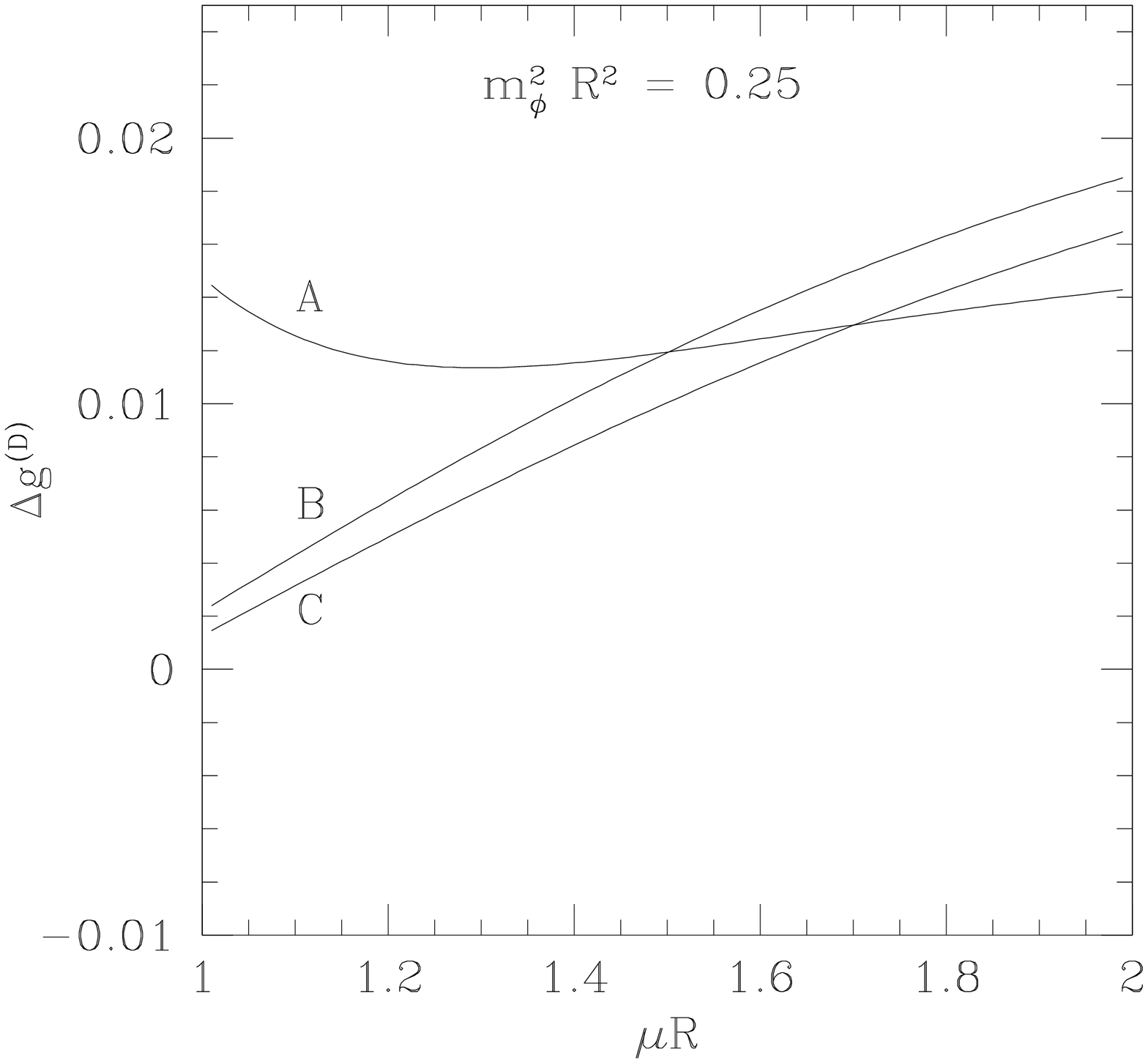} 
  \hskip 0.4 truein
   \epsfxsize 2.5 truein \epsfbox {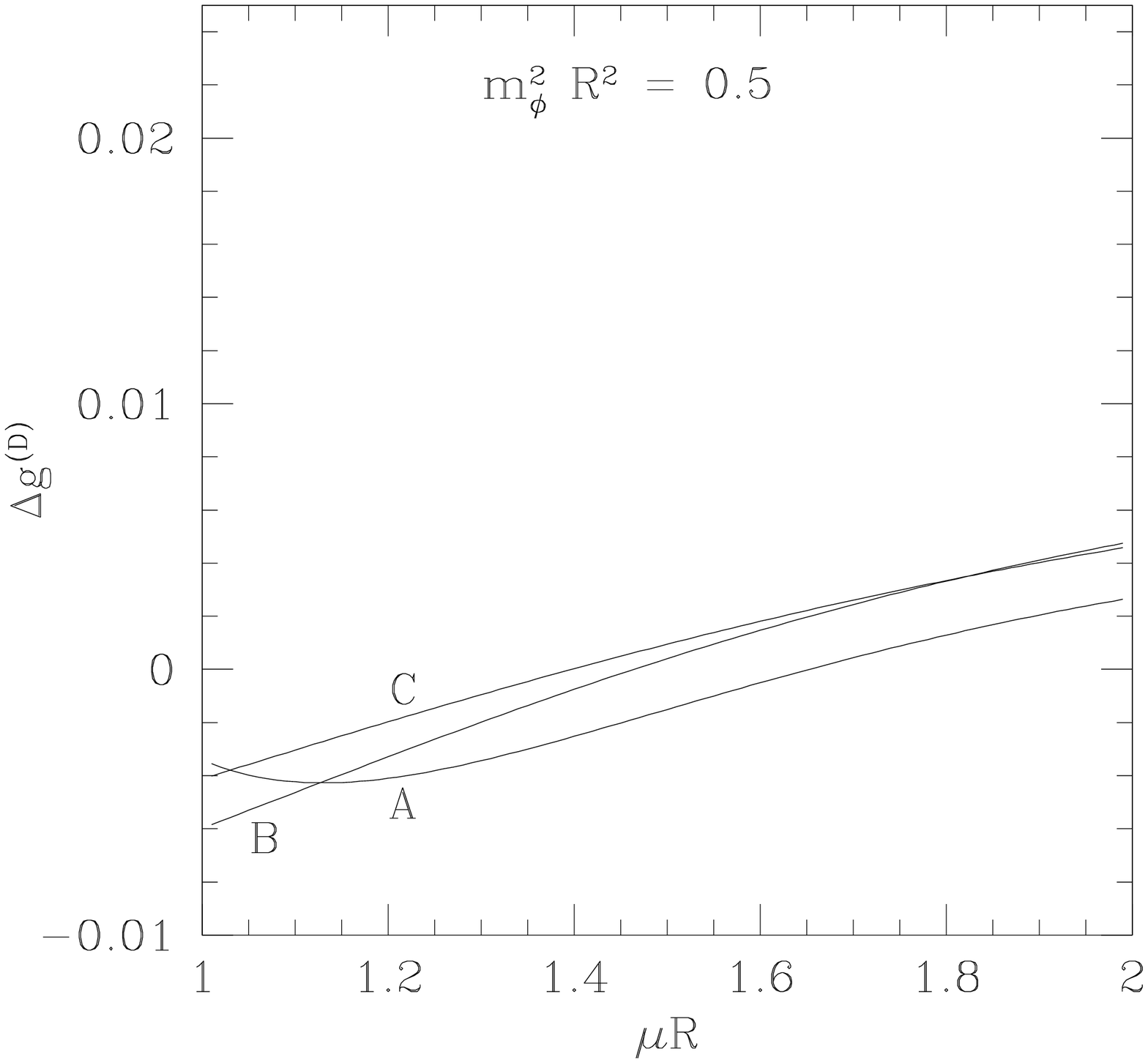} }
\vskip -0.1 truein
\caption{The total relative one-loop correction
   $L^{(D)}_{1,0} +  Z^{(\phi)}_{-1} + Z^{(\psi)}_{1} + Z^{(\psi)}_{0}$
to the coupling component $g_{n_1,0}^{(D)}$ that governs the production of the first-excited
KK fermion modes,
plotted as functions of $\mu R$ for different values of $m_\phi$ and $m_\psi$.
The plotted range in $\mu R$ extends approximately from the 
threshold for producing a particle/antiparticle pair of the first-excited KK fermion mode 
to the threshold for producing the second, assuming the $t$-channel interaction shown 
in Fig.~\protect\ref{fig11} between two incoming zero-mode fermions.
Unlike the analogous case shown in Fig.~\protect\ref{fig2b}
for the coupling in the $\lambda\phi^4$ theory, we see that the coupling that governs the production of
excited KK fermion modes in the Yukawa theory actually {\it increases}\/
relative to the zero-mode coupling 
as a function of the energy scale.  Thus, the production of excited KK fermions 
is actually slightly {\it enhanced}\/ to one-loop
order in this theory.}
\label{fig10}
\bigskip
\bigskip
\centerline{
   \epsfxsize 2.0 truein \epsfbox {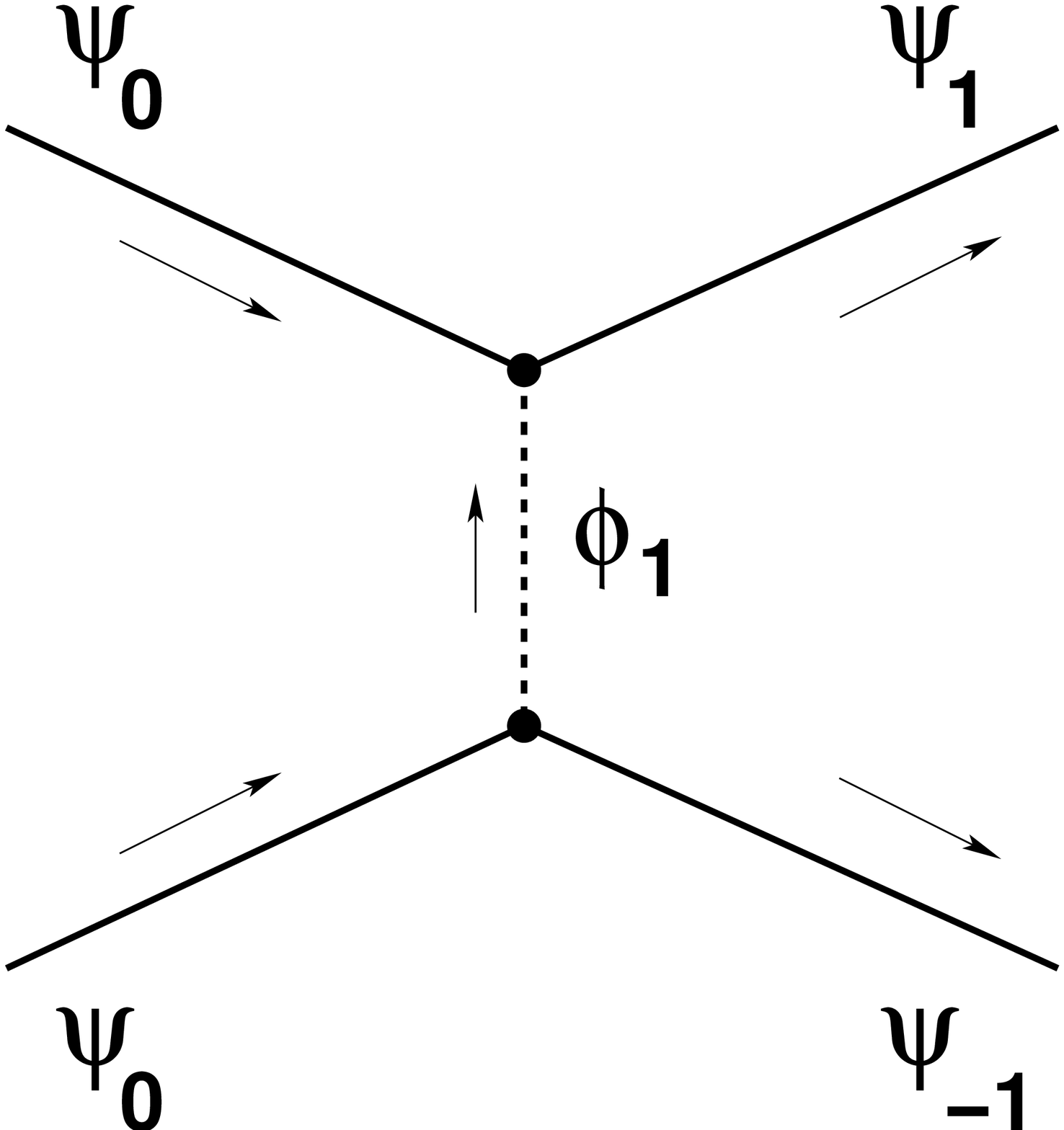} }
\caption{Feynman diagram showing the tree-level production of two first-excited KK fermions $\psi_{\pm 1}$
      via $t$-channel scattering of two zero-mode fermions $\psi_0$ in the Yukawa theory.    
      Arrows indicate the routings of five-momentum according to which the KK indices indicated
      in this figure are assigned.
      The amplitude of this process is proportional to  $g_{0,1}g_{0,-1} = g_{1,0}^2$.}
\label{fig11}
\end{figure}

In Fig.~\ref{fig10} we plot the energy dependence of the total one-loop correction
to the coupling component $g_{1,0}^{(D)}$ which governs the production of a pair of first-excited
KK fermion modes via the $t$-channel interaction 
shown in Fig.~\ref{fig11} between two incoming zero-mode fermions.
Note, in this connection, that $g_{0,1}^{(D)}=g_{0,-1}^{(D)}=g_{1,0}^{(D)}=g_{-1,0}^{(D)}$.
Relative to
the corrections to the zero-mode couplings, these KK-production couplings can be either positive or
negative, depending on the energy scale and the values of the bare
masses. As a result, we see that these one-loop corrections can either enhance or suppress
the amplitude for the creation of the first excited KK mode.
However, unlike the analogous case shown in Fig.~\ref{fig2b}
for the coupling in the $\lambda\phi^4$ theory, the coupling that governs the production of
excited KK fermion modes in the Yukawa theory actually {\it increases}\/
relative to the zero-mode coupling 
as a function of the energy scale.

Results in Fig.~\ref{fig10} are plotted for $(m_\phi R)^2=0.25$ 
and $(m_\phi R)^2=0.5$.
However, when $m_\phi =0$, 
there are
infrared divergences in the one-loop diagrams responsible for
corrections to the zero-mode coupling.  For this reason no results are plotted
in this case.  Needless to say, this is not an inconsistency:  
infrared divergences always cancel in
calculations of observables, and will do so in this higher-dimensional Yukawa
theory as well.  Indeed, such infrared
divergences also appear 
in the one-loop diagrams in the 4D Yukawa theory, and even in the 
case of full four-dimensional QED which this Yukawa theory is meant to resemble.

\begin{figure}[b!]
\vskip -0.1truein
\centerline{ \epsfxsize 2.6 truein \epsfbox {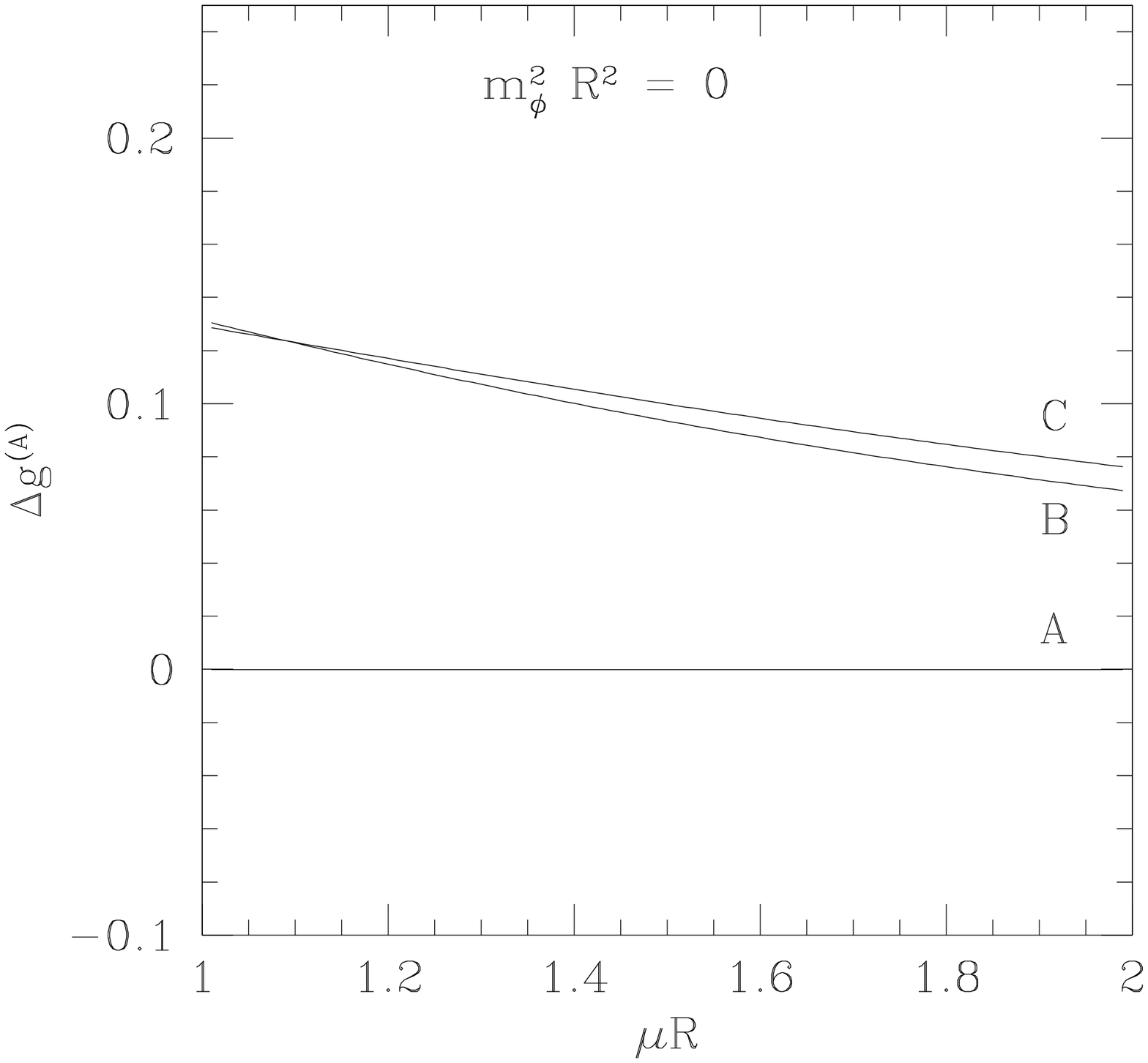}
             \hskip 0.5 truein \epsfxsize 2.6 truein \epsfbox {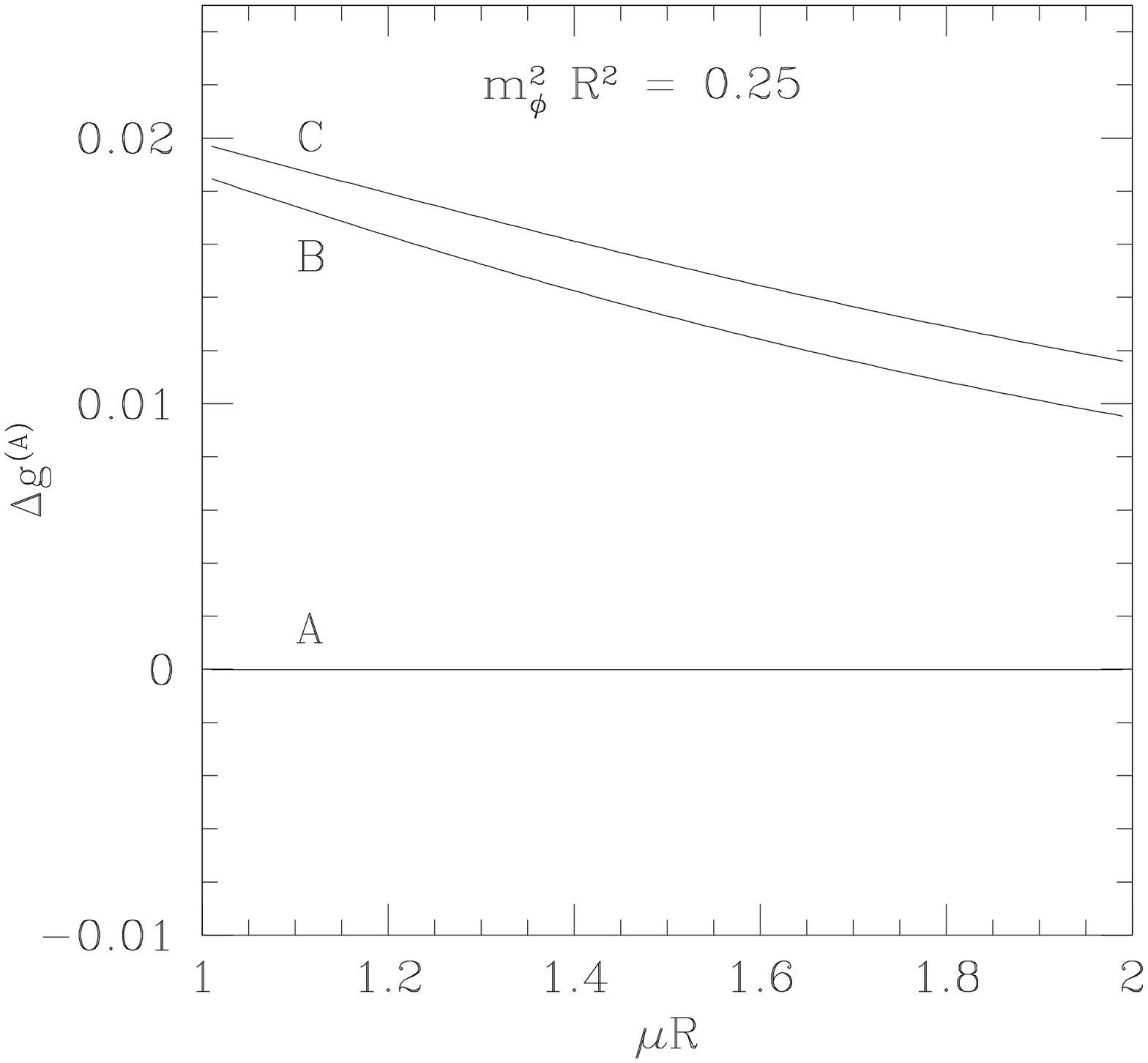} }
\vskip -0.1truein
\centerline{ \epsfxsize 2.6 truein \epsfbox {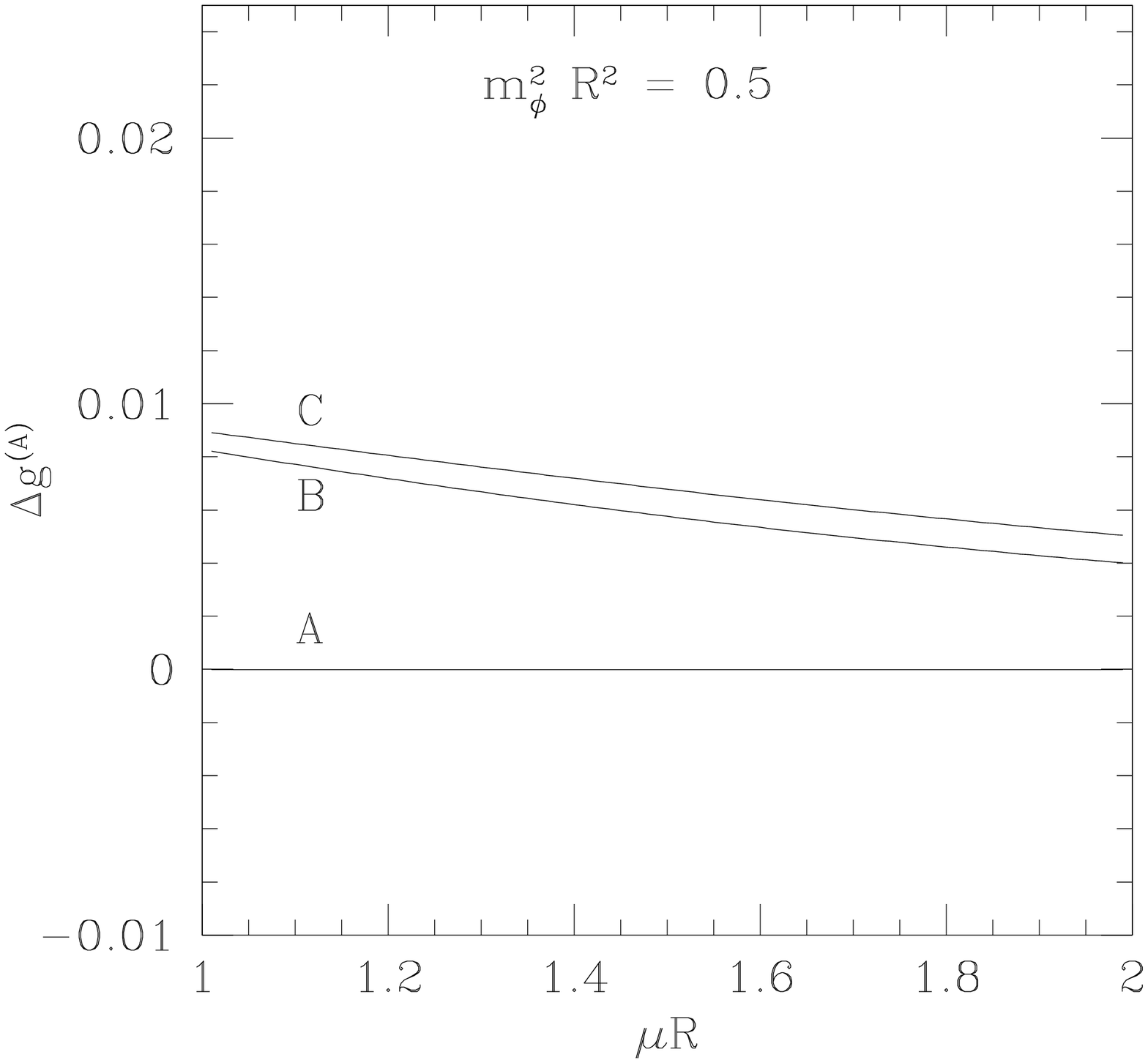} }
\vskip -0.1truein
\caption{The total relative one-loop contribution
   $L^{(A)}_{1,0}$   to the axial coupling component $g_{n_1,0}^{(A)}$,
plotted as a function of $\mu R$ for different values of $m_\phi$ and $m_\psi$.
As in Fig.~\protect\ref{fig10}, the plotted range in $\mu R$ extends approximately from the
threshold for producing a particle/antiparticle pair of the first-excited KK fermion mode
to the threshold for producing the second, assuming a $t$-channel interaction between
two incoming zero-mode fermions.}
\label{fig12}
\end{figure}

The results shown in Fig.~\ref{fig10} illustrate the Dirac-component coupling $g_{1,0}^{(D)}$.
By contrast, the corrections to the corresponding
axial coupling $g_{1,0}^{(A)}$ are shown in Fig.~\ref{fig12}. 
Unlike the Dirac coupling,
we observe that the axial coupling vanishes when $m_\psi=0$;
thus, as expected, it is the presence of non-zero $m_\psi$ which triggers a non-zero
axial coupling at one-loop order.
We also observe that this axial coupling increases monotonically as a function of $m_\psi$,
although it {\it decreases}\/ monotonically as a function of $m_\phi$. 
Furthermore, this coupling is a monotonically decreasing function of the energy
scale $\mu$;
thus, just as in the case of
the $\lambda\phi^4$ theory, the maximum coupling correction actually occurs at the threshold for
the production of the first-excited KK fermion mode. 

It is important to recognize that to one-loop order, 
the ``corrections'' shown in Fig.~\ref{fig12} are nothing but the axial couplings themselves,
since all of these axial couplings vanish at tree level.
This is therefore an instance in which a one-loop correction, though small, is
actually {\it dominant}\/.
As a consequence, any process which proceeds through such
an axial coupling is a direct probe of the 
one-loop radiative corrections we have calculated. 
Such a process, though suppressed, would be uniquely characterized
through an axial correlation between the spin and the corresponding
angular scattering amplitude.

\section{Conclusions and relation to prior work
\label{conclude2} }
\setcounter{footnote}{0}

In this paper, 
we investigated the extent
to which radiative corrections
deform the expected 
tree-level relations
between Kaluza-Klein
masses and couplings in higher-dimensional interacting theories.
Such calculations are surprisingly subtle because they rely
intrinsically on having quantum field-theoretic regulators which
preserve higher-dimensional Lorentz invariance (and higher-dimensional gauge invariance,
when appropriate);  
otherwise the standard renormalization calculations would produce spurious, unphysical effects which would
be difficult to disentangle from the {\it bona-fide}\/ physical effects resulting spacetime
compactification.  
Using techniques developed in Refs.~\cite{1a,1b}, we concentrated on two
toy theories:  five-dimensional $\lambda \phi^4$ theory and five-dimensional
Yukawa theory, each with a single dimension compactified on a circle.
We then studied the resulting one-loop corrections to the tree-level
mass and coupling relations, and determined those situations in which 
these corrections exhibited a variety of special algebraic forms and behaviors
as functions of the bare five-dimensional masses in these theories
and the overall renormalization
energy scale.

For both $\lambda\phi^4$ theory and Yukawa theory on a circle,
we found that our KK masses can deform in a variety of different ways.
In some cases, these deformations do not disturb the underlying
KK mass {\it relations}\/ between different KK modes.   
In such cases, therefore, the underlying five-dimensional Lorentz invariance
of the KK mass spectrum appears to be preserved.
In other cases, these deformations induce changes in these relations 
which can be interpreted as mere shifts or ``renormalizations'' of
the underlying five-dimensional masses or the radius of the compactification
circle.  However, in the most general cases, these deformations result
in new KK mass relations which do not exhibit the signatures normally
associated with compactification on a circle. 

Similar results were also found for the KK couplings:
renormalization effects can induce
non-trivial splittings between KK couplings 
which are otherwise equal at tree level.
For $\lambda \phi^4$ theory, we found that these splittings
lead to enhanced production of the first-excited KK mode.
In Yukawa theory, by contrast, we found that renormalization effects
can lead to either enhanced
or suppressed production of the first-excited KK mode.  
Whether this production is ultimately enhanced or suppressed depends on the values of 
the underlying five-dimensional masses and the
energy scale of the experiment through which it is measured. 

While many of our results were expected,
others were more surprising.
One interesting result, for example, is the radiative generation of
a $\gamma^5$-interaction amongst zero modes in the Yukawa theory.  
Indeed, such an interaction is completely absent at tree level.
As we discussed in Sect.~4, this interaction does not
lead to parity or CP violation, and is analogous to the axial fermion mass terms which
appear in the KK Lagrangian at tree level.
Another somewhat surprising result 
is that the corrections to the
axial masses of the fermions in Yukawa theory vanish when the
zero-mode masses of the boson and fermion are equal.  As we briefly discussed in 
Sect.~4,
this cancellation ultimately occurs
because the one-loop corrections to fermion propagators in Yukawa theory
are equivalent to those in a supersymmetric model, up to an overall
multiplicative constant. Supersymmetry should forbid axial mass
corrections.

Needless to say,
many previous studies have focused on loop corrections in KK theories. 
However, most of this prior work focused on the effects induced by
the excited KK states on the properties of the zero modes. 
For example, a relatively early calculation of the runnings of zero-mode
gauge couplings appears in Ref.~\cite{DDG},
where it was found that the higher-dimensional
radiative corrections to such runnings have the potential to lead to gauge coupling 
unification well below the usual GUT scale.  Such running can also generate 
fermion mass hierarchies~\cite{DDG}.
However, the analysis of Ref.~\cite{DDG} focused purely on the radiative
corrections to the couplings of the zero modes, and thus did not require
use of regulators designed to respect higher-dimensional Lorentz or gauge symmetries. 
Likewise, the authors of Ref.~\cite{pomarol} calculated gauge-coupling
corrections in warped AdS$_5$ space.  A recent study of loop
effects in this geometry appears in Ref.~\cite{choi_kim_shin}.

Another type of zero-mode calculation involves the special case in
which loop corrections are finite to a certain order in perturbation
theory. This variety of calculation appears in Ref.~\cite{muon}, for example, where
the authors calculated the correction to the muon magnetic
moment in higher dimensions.  At one-loop order, the correction was
found to be finite in 5D.

There do, however, exist several studies which have examined 
loop effects on excited modes. For example, the authors
of Ref.~\cite{GGH} showed that when an extra dimension is compactified
to an orbifold, loop corrections lead to logarithmically divergent
terms localized at the orbifold fixed points. These can take the form
of new kinetic terms or coupling terms at the fixed points.

The authors of Ref.~\cite{CMS1} calculated corrections to KK masses in
five-dimensional QED and in a five-dimensional Standard Model, considering
the cases in which these theories are compactified 
on a flat, circular universal extra dimension and on a flat $S^1/\IZ_2$ orbifold.
For the
case of compactification on a circle, they found
that if zero-mode fermion masses are
neglected, the photon zero mode remains massless while the excited KK photons receive 
mass corrections of the form
\beq
\Delta m_{n}^{2} ~=~ -\frac{e^2 \zeta(3)}{4\pi^4 R^2}~,~~~~n\geq 1~,
\label{KK_photon}
\eeq
independent of the mode number $n$.  
Other gauge theories lead to similar
results. 
This sort of behavior is clearly an example of Case~\#3a from the Introduction:
the entire excited tower experiences a uniform mass shift, while 
gauge invariance protects the (vanishing) mass of the gauge-boson zero mode.

The authors of Ref.~\cite{CMS1} correctly obtained this result by
performing a Poisson resummation, casting KK sums into sums over winding numbers.
Indeed, the use of
Poisson resummations in calculations of loop corrections first
appeared in Ref.~\cite{antoniadis}, and it has been verified~\cite{1a} that use of
our regulators also reproduces the result in Eq.~(\ref{KK_photon}).    
At first glance, it might seem that
such a Poisson-resummation technique might also apply to the calculations
in this paper. Unfortunately, this is not the case because this method does not yield
closed-form expressions when the zero-mode masses are non-vanishing.  Indeed,
as we have seen, many of our results arise precisely because of the non-vanishing
nature of these masses.
As a result, regulators of the type we introduced in Refs.~\cite{1a} and~\cite{1b}
are needed for the calculations in this paper.

As an aside, we remark that
there also remains the
technical issue that a Poisson resummation by itself does not
regularize a divergence, but merely expresses it in a different language.  
In Ref.~\cite{1a}, for example, we noted that Poisson
resummation
worked in Ref.~\cite{CMS1} because the mass corrections
in those calculations
were finite.  For the divergent case, however, we noted that one would have to
calculate {\it differences}\/ between corrections for excited modes and
zero modes, analogous to the differences introduced in
Refs.~\cite{1a} and~\cite{1b}.  Of course, one might be tempted to simply subtract
the contribution arising from vanishing winding number.
However, this merely corresponds to the correction in a non-compactified theory, and does
not relate directly to observables in the compactified theory.

The authors of Ref.~\cite{mixed} calculated loop corrections to KK 
gauge-boson masses using a mixed propagator. In this approach, 
the four large dimensions are treated
in momentum space, as usual, while the compactified extra
dimension is treated in position space.  This avoids the
introduction of a KK sum altogether.
However, in such situations the higher-dimensional divergences
are not eliminated ---
they are the same as would appear in the corresponding
higher-dimensional {\it uncompactified}\/ theory, as this formalism
makes abundantly clear.
Of course, it is possible that the 
true UV limit of a given higher-dimensional
theory is not higher-dimensional at all~\cite{deconstruction}.
Such ``deconstructed'' extra dimensions would change the UV divergence structure
of the theory in a profound way that would eliminate the need for many of these
different regularization techniques, and indeed it has been demonstrated~\cite{deconstruct}
that such deconstruction techniques lead to results which are 
consistent with those in Ref.~\cite{CMS1}
and in other papers.

In a similar vein, radiative corrections may be finite in cases in which
there exist additional symmetries (either unbroken or softly broken)
to protect against divergences. 
Well-known examples of this phenomenon include radiative corrections in theories with supersymmetry
broken through the Scherk-Schwarz mechanism~\cite{SS},
or in theories
in which the Higgs is identified as a component of a higher-dimensional
gauge field and consequently has a mass for which radiative corrections
are protected by gauge symmetries~\cite{hosotani}.

The authors of Ref.~\cite{gauge_higgs1} calculated loop corrections to
the KK masses of gauge bosons in a theory with an extra dimension compactified on an
$S^1/\IZ_2$ orbifold.  Like the authors of Ref.~\cite{CMS1}, they used
Poisson resummation techniques to calculate bulk effects and the methods
of Ref.~\cite{GGH} to calculate brane terms. 
By explicitly calculating loop diagrams, they
showed that quadratic divergences to the Higgs mass are avoided. This
is in agreement with a previous study~\cite{gauge_higgs2}, which
showed that quadratic divergences are avoided in a particular model
involving gauge-Higgs unification. 
However, these analyses take place within the contexts of
theories in which a higher-dimensional gauge theory is broken to a gauge
subgroup at orbifold fixed points via the Hosotani mechanism~\cite{hosotani}.

Another approach to loop corrections in higher dimensions is to embed
Kaluza-Klein theory into string theory, and to perform string-theory
calculations. Indeed, the authors of Ref.~\cite{DKR_string} analyzed 
higher-dimensional vacuum polarization diagrams in this context, and
reproduced the gauge-boson KK mass corrections discussed above. 
This correspondence holds when the string
scale is much greater than the inverse radius of the extra
dimension. 
Other string-motivated methods of dealing with the divergences in higher-dimensional
theories are discussed in Refs.~\cite{DMM_string,GNS_string}.
In a similar vein, the authors of Ref.~\cite{DKR_string} demonstrated
that similar results can be obtained using techniques from lattice field theory. 
Of course, this assumes that 
the lattice spacing is much smaller than the compactification radius.
Other regularization techniques for KK theories are discussed 
in Refs.~\cite{contino_pilo,cas_dimreg,alvarez_faedo}. 

Quantum corrections involving KK states are also relevant to the 
calculations of Casimir energies, and more generally to the evaluation
of the {\it stability}\/ of an extra dimension. 
As a result,  there have been a number of papers examining 
topics along these lines.
For example, the authors of
Ref.~\cite{cas1} examined a gravitational analogue of the Casimir
effect along an extra dimension compactified on a circle
using a hard cutoff to regularize momenta of KK states.
Other techniques have also been used~\cite{cas2,cas_ferm,witten,gold_wise_roth,ichinose}.

Finally, we remark that in addition to quantum corrections in higher-dimensional theories,
there are also non-trivial {\it classical}\/ effects which can also distort the ``apparent''
geometry of an extra dimension as measured through analyses of KK spectroscopy.
Indeed, the authors of
Ref.~\cite{classical} showed that the geometry of an extra dimension
can even experience a type of classical renormalization.

Needless to say, there are a number of extensions
of this work that may be pursued in the future.
For example, one avenue is to calculate radiative corrections in 
higher-dimensional theories with supersymmetry.  Such an analysis may permit a determination
of what radiative effects are allowed in supersymmetric models, and how
the radiative effects on KK bosons and fermions come into alignment.
This question is particularly relevant in our case, since we have already seen
that the axial mass corrections in the non-supersymmetric Yukawa theory analyzed here 
actually vanish in a limit corresponding to supersymmetry. 
Another avenue for future research is to 
employ the regulators developed in
Refs.~\cite{1a} and~\cite{1b} in order to analyze decays of KK 
modes in higher-dimensional theories;  indeed, preliminary results along these lines~\cite{2b}
suggest a number of striking properties which may have deep significance for 
the phenomenological properties and ultimately the stability properties of these modes.
This may be particularly relevant for recent discussions of dynamical dark matter~\cite{DDM}.
Finally, a third avenue for further research involves an examination of 
more realistic compactification scenarios, especially those involving orbifolds 
(rather than manifolds), as needed in order to produce chiral four-dimensional theories.
Work along all of these lines is in progress.

\bigskip

\section*{Acknowledgments}
\setcounter{footnote}{0}

We are happy to thank Z.~Chacko, S.~Su, D.~Toussaint, and U.~van~Kolck for discussions.
This work is supported in part by the Department of Energy
under Grants~DE-FG02-04ER-41298 and DE-FG02-08ER-41531, and by the Wisconsin
Research Alumni Foundation.
The opinions and conclusions expressed here are those of the authors,
and do not represent either the Department of Energy or the National Science Foundation.

\bigskip

\bibliographystyle{unsrt}

\end{document}